\documentclass[12pt]{article}
\usepackage[dvips]{graphicx}
\usepackage{latexsym}
\parskip=\medskipamount

\def \un{\underline}
\newcommand {\cD}{{\cal D}}

\newcommand {\cM}{{\cal M}}
\newcommand {\cN}{{\cal N}}

\newcommand {\cQ}{{\cal Q}}

\newcommand {\cT}{{\cal T}}

\newcommand {\cW}{{\cal W}}

\newcommand{\bG}{{\bf G}}

\newcommand{\bK}{{\bf K}}

\newcommand{\bW}{{\bf W}}

\def\a{\alpha}
\def \bi{\bibitem}

\def\b{\beta}

\def\d{\delta}

\def\f{\phi}

\def\G{\Gamma}

\def\l{\lambda}
\def\m{\mu}
\def\n{\nu}

\def\p{\pi}
\def\q{\theta}
\def\r{\rho}
\def\s{\sigma}
\def\t{\tau}

\def\x{\xi}
\def\z{\zeta}

\def\F{\Phi}
\def\J{\Psi}

\def\S{\Sigma}
\def\U{\Upsilon}

\newcommand{\ad}{{\dot{\alpha}}}                           
\newcommand{\bd}{{\dot{\beta}}}                            
\newcommand{\ve}{\varepsilon}                            

\newcommand{\hf}{\frac12}

\newcommand{\vf}{\varphi}
\newcommand{\sect}[1]{\setcounter{equation}{0}\section{#1}}

\newcommand{\be}{\begin{equation}}
\newcommand{\ee}{\end{equation}}
\newcommand{\bea}{\begin{eqnarray}}
\newcommand{\eea}{\end{eqnarray}}
\newcommand{\non}{\nonumber}
\begin{document}

\begin{titlepage}
\thispagestyle{empty}

\begin{flushright}
hep-th/0310025 \\
October, 2003 \\
\end{flushright}
\vspace{5mm}

\begin{center}
{\Large \bf
On the two-loop four-derivative quantum corrections\\
in 4D N = 2 superconformal field theories}
\end{center}

\begin{center}
{\large S. M. Kuzenko and   I. N. McArthur}\\
\vspace{2mm}

\footnotesize{
{\it School of Physics, The University of Western Australia\\
Crawley, W.A. 6009, Australia}
} \\
{\tt  kuzenko@cyllene.uwa.edu.au},~
{\tt mcarthur@physics.uwa.edu.au} \\
\end{center}
\vspace{5mm}

\begin{abstract}
\baselineskip=14pt
In  $\cN=2, 4$ superconformal field theories in 
four space-time dimensions, 
the quantum corrections with four derivatives are believed 
to be severely constrained by non-renormalization theorems.  
The strongest of these is the conjecture formulated 
by Dine and Seiberg  in hep-th/9705057 that such terms 
are generated only at one loop. In this note, using 
the background field formulation 
in $\cN=1$ superspace,  we test the Dine-Seiberg proposal 
by comparing  the two-loop $F^4$ quantum corrections 
in two different superconformal 
theories with the same gauge group $SU(N)$: 
(i) $\cN=4$ SYM (i.e. $\cN=2$ SYM with a single adjoint 
hypermultiplet); (ii) $\cN=2$ SYM with $2N$ hypermultiplets
in the fundamental. According to the Dine-Seiberg conjecture, 
these theories should yield identical two-loop $F^4$ contributions
from all the supergraphs  involving quantum hypermultiplets, 
since the pure $\cN=2$ SYM and ghost sectors are identical
provided the same gauge conditions are chosen.
We explicitly evaluate the relevant two-loop  supergraphs and 
observe that the $F^4$ corrections generated  
have different large $N$ behaviour in the two theories under 
consideration. Our results are in  conflict 
with the Dine-Seiberg conjecture. 
\end{abstract}

\vfill
\end{titlepage}

\newpage
\setcounter{page}{1}

\renewcommand{\thefootnote}{\arabic{footnote}}
\setcounter{footnote}{0}
\sect{Introduction}
Some time ago, we developed 
a manifestly covariant approach 
for evaluating multi-loop quantum corrections
to low-energy effective actions within the 
background field formulation \cite{KM}. 
This approach is  applicable to ordinary gauge theories
and  to supersymmetric Yang-Mills theories
formulated in superspace. Its power is not restricted
to computing just the counterterms --  it is well suited
for deriving finite quantum corrections in 
the framework of the derivative 
expansion. 
As a simple application of the techniques 
developed in \cite{KM}, 
we have recently derived  \cite{KM2}
the two-loop (Euler-Heisenberg-type) 
effective action for $\cN=2$ supersymmetric QED 
formulated in  $\cN=1$ superspace. 

The work of \cite{KM2} has brought a surprising outcome
regarding one particular conclusion drawn 
in \cite{BKO} on the basis of the background field 
formulation in $\cN=2$ harmonic superspace \cite{BBKO}.
According to \cite{BKO},  no super $F^4$ (four-derivative) 
quantum corrections occur at two loops in generic $\cN=2$ 
super Yang-Mills theories on the Coulomb branch, 
in particular in $\cN=2$ SQED.
However, by explicit calculation carried out in \cite{KM2}, 
it was shown that  a non-vanishing 
two-loop $F^4$ correction does occur in $\cN=2$ SQED. 
It was also shown in \cite{KM2}
that  the analysis in  \cite{BKO} 
contained a subtle loophole related
to the intricate structure of harmonic supergraphs. 
A more careful treatment of two-loop harmonic supergraphs
given in \cite{KM2} leads to the same non-zero $F^4$  term 
 in $\cN=2$ SQED at two loops as that derived 
using the $\cN=1$ superfield formalism.

The work of \cite{BKO} provided perturbative two-loop 
support  to the famous Dine-Seiberg 
non-renormalization conjecture
\cite{DS} that the 
$\cN=2$ supersymmetric 
four-derivative term\footnote{The 
functional (\ref{N=2F4}) was originally introduced 
in \cite{dWGR}. It is a unique $\cN=2$ superconformal 
invariant  in the family of non-holomorphic actions 
of the form $\int {\rm d}^4 x {\rm d}^8 \q \,H(W, {\bar W})$
introduced for the first time in \cite{Hen}.
More general (higher-derivative) superconformal 
invariants of the  $\cN=2$ Abelian vector multiplet   
 were given in \cite{BKT}.} 
\bea
 \int {\rm d}^4 x {\rm d}^8 \q \,
\ln W \,\ln {\bar W} 
\label{N=2F4}
\eea
is one-loop exact on the Coulomb branch of  $\cN=2,4$ 
superconformal theories.\footnote{The one-loop 
$F^4$ quantum corrections in $\cN=2,4$ 
SQFTs were computed in 
\cite{GR,BK2,LvU}.} 
It is known that the Dine-Seiberg  conjecture is  
well supported by non-perturbative considerations
\cite{DKMSW,BFMT}.
But since the two-loop $F^4$ conclusion of \cite{BKO}
is no longer valid, it seems important 
to carry out  an  independent calculation of the two-loop $F^4$ 
quantum corrections in $\cN=2$ superconformal theories.  
It is the aim of the present note to provide 
such a calculation.  As will be demonstrated below, 
the Dine-Seiberg  conjecture
is not fully supported  at the  perturbative level.

To test the Dine-Seiberg conjecture,
we consider 
two different $\cN=2$ superconformal 
theories with the same gauge group $SU(N)$: 
(i) $\cN=4$ SYM or, equivalently,  
$\cN=2$ SYM with a single adjoint hypermultiplet; 
(ii) $\cN=2$ SYM with $2N$ hypermultiplets
in the fundamental. At the quantum level, 
with the same gauge conditions chosen, 
these theories are identical in the pure $\cN=2$ SYM 
and ghost sectors. The difference between them 
occurs only in the sector involving quantum 
hypermultiplets. If the Dine-Seiberg conjecture holds, 
then since the pure $\cN=2$ SYM 
and ghost sectors are identical, 
these theories should yield identical two-loop $F^4$ contributions
from all the supergraphs  with quantum hypermultiplets.
However, as will be shown below by  direct calculations,
the relevant two-loop $F^4$ contributions have 
different large $N$ behaviour in the theories 
under  consideration.\footnote{To test the Dine-Seiberg 
conjecture, we do not need the two-loop $F^4$ contribution
from the pure $\cN=2$ SYM 
and ghost sectors. It will be discussed in a separate paper.}  

 ${}$From the point of view of $\cN=1$ supersymmetry,
the chiral superfield strength $W$ of the 
$\cN=2$ vector multiplet  is known 
to consist of a chiral  scalar $\f$ and a constrained chiral 
spinor $W_\a$, the latter being 
the $\cN=1$ vector multiplet field strength. 
When reduced to $\cN=1$ superspace, 
the functional (\ref{N=2F4}) is given by a sum of several terms, 
of which the leading (in a derivative expansion) term is 
\be
\U = \int {\rm d}^8 z \, \frac{ W^\a W_\a
{\bar W}_\ad {\bar W}^\ad  }{\f^2{\bar \f}^2 }~,
\label{F4}
\ee 
while the other terms involve derivatives of $\f$ and $\bar \f$.
If one uses the $\cN=1$ superspace formulation 
for $\cN=2$ superconformal field theories, it is 
typically sufficient to compute quantum corrections 
of the form (\ref{F4})  in order to restore their
$\cN=2$ completion (\ref{N=2F4}).

This note is organized as follows. 
Section 2 contains the necessary setup
regarding $\cN=2$ superconformal 
field theories and their background field
quantization (for supersymmetric 't Hooft gauge)
 in $\cN=1$ superspace. In section 3 we work out a useful 
functional representation for two-loop supergraphs
with quantum hypermultiplets.
In section 4 we describe,  following \cite{KM,KM2}, 
the exact superpropagators in a special $\cN=2$ 
vector multiplet background which is 
extremely simple but perfectly suitable for
computing quantum corrections of the form (\ref{F4}).
Sections 5 and 6 form the (somewhat technical) core of this paper. 
In section 5 we evaluate the two-loop $F^4$ corrections
in $\cN=2$ SYM with $2N$ hypermultiplets
in the fundamental. This consideration is extended in section 
6 to the case of $\cN=4$ SYM. 
Finally, in section 7 we compare the two-loop corrections 
in the large $N$ limit for the two theories being 
studied.  Some aspects of the cancellation of divergences 
are  discussed  in the appendix.

\sect{\mbox{$\cN = 2$} SYM setup}

The classical action of an $\cN=2$ superconformal 
field theory, 
$S_{\rm SCFT} = S_{\rm vector} + S_{\rm hyper}$,
consists of two parts: (i) the pure $\cN=2$ SYM action 
\bea
S_{\rm vector} &=& \frac{1}{g^2}\,{\rm tr}_{\rm F}
\Big( \int {\rm d}^8 z \, \F^\dagger \F
+  \int {\rm d}^6 z \, \cW^\a \cW_\a 
\Big)~;
\label{n=2pure-sym}
\eea
 (ii) the hypermultiplet action
\bea 
S_{\rm hyper} = 
 \int {\rm d}^8 z \, \Big( \cQ^\dagger  \,\cQ 
+  \tilde{\cQ}^\dagger   \tilde{\cQ} \Big)
 - {\rm i}  \int {\rm d}^6 z \, \tilde{\cQ}^{\rm T} \F \cQ 
+ {\rm i}  \int {\rm d}^6 {\bar z}\,  \cQ^\dagger 
\F^\dagger \overline{\tilde{\cQ}} ~.
\label{hyper}
\eea
Here $\F$, $Q$ and $\tilde{Q}$ are {\it covariantly chiral}
superfields which transform, respectively, in the following 
representations of the gauge group:
(1) the adjoint; (2) a  representation $\rm R$; and
(3) its conjugate  ${\rm R}_{\rm c}$. 
The covariantly chiral superfield strength $\cW_\a$ 
is associated with  the gauge  covariant derivatives
\be
\cD_A = (\cD_a, \cD_\a , {\bar \cD}^\ad ) 
= D_A +{\rm i}\, \G_A~, \qquad 
\G_A = \G^\m_A (z) T_\m~, \qquad 
(T_\m)^\dagger = T_\m~, 
\ee
where $D_A$ are the flat covariant 
derivatives\footnote{Our $\cN=1$ notation 
and conventions correspond to \cite{BK}.}, 
and $\G_A$ the superfield connection taking its values 
in the Lie algebra of the gauge group.  
The   gauge covariant derivatives satisfy the following algebra:
\bea
& \{ \cD_\a , \cD_\b \} 
= \{ {\bar \cD}_\ad , {\bar \cD}_\bd \} =0~, \qquad 
\{ \cD_\a , {\bar \cD}_\bd \} = - 2{\rm i} \, \cD_{\a \bd}~, \non \\
& [ \cD_\a , \cD_{\b \bd}] = 2 {\rm i} \ve_{\a \b}\,{\bar \cW}_\bd ~, 
\qquad 
[{\bar \cD}_\ad , \cD_{\b \bd}] = 2{\rm i} \ve_{\ad \bd}\,\cW_\b ~ , 
\non \\
& [ \cD_{\a \ad}, \cD_{\b \bd} ] 
= - \ve_{\a \b}\, {\bar \cD}_\ad {\bar \cW}_\bd 
-\ve_{\ad \bd} \,\cD_\a \cW_\b~. 
\label{N=1cov-der-al}
\eea
The spinor field strengths $\cW_\a$ and ${\bar \cW}_\ad$ 
obey the Bianchi  identities
\be
{\bar \cD}_\ad \cW_\a =0~, \qquad 
\cD^\a \cW_\a = {\bar \cD}_\ad {\bar \cW}^\ad~.
\ee
The condition under which the $\cN=2$ theory 
is finite is  
\be
{\rm tr}_{\rm Ad} \,\F^2 ~=~ {\rm tr}_{\rm R} \, \F^2~.
\label{finita}
\ee
It is assumed that in the action (\ref{n=2pure-sym}) 
 the superfields $\F$ and $\cW_\a$ 
are given in the fundamental (or defining) 
representation of the gauge group, 
with  the corresponding generators
normalized such that  
${\rm tr}_{\rm F} \,(T_\m\, T_\n) = \d_{\m \n}$.

To quantize the theory, we will use the $\cN=1$ 
background field formulation \cite{GGRS} and 
split the dynamical variables into background 
and quantum ones, 
\bea
 \F ~ \to ~ \F +\vf ~, \qquad \cQ ~ &\to & ~ \cQ +q ~, 
\qquad 
\tilde{\cQ} ~ \to ~ \tilde{\cQ}+ \tilde{q} ~, \non \\
  \cD_\a ~ \to ~ {\rm e}^{-v} \, \cD_\a \, {\rm e}^v~, 
\quad && \quad 
{\bar \cD}_\ad ~ \to ~ {\bar \cD}_\ad~,
\eea
with lower-case letters used for 
the quantum superfields. 
In this paper, we are not interested in the 
dependence of the effective action on 
the hypermultiplet superfields, 
and therefore we set $\cQ = \tilde{\cQ} =0$
in what follows. 
After the background-quantum splitting, 
the action (\ref{n=2pure-sym}) turns into
\bea
S_{\rm vector} &=& \frac{1}{g^2}\,{\rm tr}_{\rm F}
\Big( \int {\rm d}^8 z \, (\F +\vf)^\dagger \,
{\rm e}^v \, (\F +\vf) \, {\rm e}^{-v}
+  \int {\rm d}^6 z \, \bW^\a \bW_\a 
\Big)~,
\label{bqs-vm}
\eea
where 
\bea
\bW_\a &=& - {1\over 8} {\bar \cD}^2 \Big( 
{\rm e}^{-v}\, \cD_\a \,{\rm e}^{v} \cdot 1 \Big)
= \cW_\a - {1\over 8} {\bar \cD}^2 \Big( 
\cD_\a v- \hf [v, \cD_\a v] \Big) ~+~ O(v^3)~. 
\eea 
The hypermultiplet  action (\ref{hyper}) takes the form 
\bea 
S_{\rm hyper} = 
 \int {\rm d}^8 z \, \Big( q^\dagger \, {\rm e}^v \,q 
+  \tilde{q}^\dagger \, {\rm e}^{-v^{\rm T}} \, \tilde{q} \Big)
 - {\rm i}  \int {\rm d}^6 z \, \tilde{q}^{\rm T} 
(\F + \vf)  q 
+ {\rm i}  \int {\rm d}^6 {\bar z}\,  q^\dagger 
(\F +\vf)^\dagger \overline{\tilde{q}} ~.
\label{bqs-hyper}
\eea

It is advantageous to use 
$\cN=1$ supersymmetric 't Hooft gauge 
(a special case of the supersymmetric $R_\x$-gauge
introduced in \cite{OW} and further developed in \cite{BBP})
which is specified by the nonlocal gauge conditions 
\bea
-4 \chi \,&= &{\bar \cD}^2 v +
[\F, ({\Box_+})^{-1} {\bar \cD}^2 \vf^\dagger ] 
= {\bar \cD}^2 v +
[\F, {\bar \cD}^2  ({\Box_-})^{-1}  \vf^\dagger ] 
~, \non \\
-4 \chi^\dagger  &= &\cD^2 v -
[\F^\dagger , ({\Box_-})^{-1} \cD^2 \vf ] 
= \cD^2 v -
[\F^\dagger , \cD^2  ({\Box_+})^{-1}  \vf ] ~.
\eea
Here the covariantly chiral d'Alembertian, $\Box_+$,
is defined by
\bea 
\Box_+ &=& \cD^a \cD_a - \cW^\a \cD_\a -\hf \, (\cD^\a \cW_\a)~, 
\quad
\Box_+ \J = {1 \over 16} \, {\bar \cD}^2 \cD^2 \J ~, \quad 
{\bar \cD}_\ad \J =0~,
\eea
for a covariantly chiral superfield $\J$.
Similarly, the  covariantly antichiral 
d'Alembertian, $\Box_-$, is defined by
\bea 
\Box_- &=& \cD^a \cD_a + {\bar \cW}_\ad {\bar \cD}^\ad 
+\hf \, ({\bar \cD}_\ad  {\bar \cW}^\ad)~, 
\quad
\Box_- {\bar \J} = {1 \over 16} \, \cD^2 {\bar \cD}^2  {\bar \J} ~, 
\quad  \cD_\a {\bar \J} =0~,
\eea
for a covariantly antichiral superfield $\bar \J$.
The gauge-fixing functional\footnote{In this paper, 
the explicit structure of the ghost sector is not 
required.}  is
\be
S_{\rm GF} = -\frac{1}{g^2}\,{\rm tr}_{\rm F}
 \int {\rm d}^8 z \, \chi^\dagger \,\chi~.
\ee
The quantum quadratic part of 
$S_{\rm vector} + S_{\rm GF}$ is 
\bea 
S^{(2)}_{\rm vector} + S_{\rm GF} 
 &=&\, ~ \frac{1}{g^2}\,{\rm tr}_{\rm F}
 \int {\rm d}^8 z \,
\Big( \vf^\dagger \,\vf 
- [\F^\dagger , [\F, \vf^\dagger ]] \, 
\frac{1}{\Box_+} \vf \Big) \non \\
&-&\frac{1}{2g^2}\,{\rm tr}_{\rm F} \int {\rm d}^8 z \,
v \, \Big(  \Box_{\rm v}  v -[\F^\dagger, [\F, v]] \Big) 
~+~ \dots 
\label{quad-prel}
\eea 
where the dots stand for the terms with derivatives
of the background (anti)chiral superfields 
 $\F^\dagger$ and $\F$.
The vector d'Alembertian,
$\Box_{\rm v}$,  is defined by
\bea 
{\Box}_{\rm v} 
&=& \cD^a \cD_a - \cW^\a \cD_\a +{\bar \cW}_\ad {\bar \cD}^\ad \\
&=& -\frac{1}{8} \cD^\a {\bar \cD}^2 \cD_\a 
+{1 \over 16} \{ \cD^2 , {\bar \cD}^2 \} 
-\cW^\a \cD_\a -\hf  (\cD^\a \cW_\a) \non \\
&=& 
 -\frac{1}{8} {\bar \cD}_\ad \cD^2 {\bar \cD}^\ad 
+{1 \over 16} \{ \cD^2 , {\bar \cD}^2 \} 
+{\bar \cW}_\ad {\bar \cD}^\ad +\hf({\bar \cD}_\ad {\bar \cW}^\ad ) ~.
\non 
\eea
The quantum quadratic part of $S_{\rm hyper} $ is
\bea 
S^{(2)}_{\rm hyper} = 
 \int {\rm d}^8 z \, \Big( q^\dagger  \,q 
+  \tilde{q}^\dagger  \, \tilde{q} \Big)
 +  \int {\rm d}^6 z \, \tilde{q}^{\rm T}\, \cM_{\rm R}  \,q 
+   \int {\rm d}^6 {\bar z}\,  q^\dagger \,
\cM_{\rm R}^\dagger \, \overline{\tilde{q}} ~.
\label{hyper-quad}
\eea
Here the operator $\cM$ is defined by 
\bea
\cM_{\rm D} \,\S &=& -{\rm i}\, \F\, \S~,
\eea
for a superfield $\S$ transforming 
in some representation D of the gauge group.

The background superfields will be  chosen to form 
a special on-shell $\cN=2$ vector multiplet in the Cartan 
subalgebra of the gauge group:
\be
[\F , {\bar \F} ] ~= ~ \cD^\a \cW_\a = 0~, 
\qquad \cD_\a \F ~=~0~.
\ee
Such a background configuration is convenient 
for computing those corrections to the effective action 
which do not contain derivatives of $\F$ and $\F^\dagger$.
Now, the action (\ref{quad-prel}) becomes
\bea 
S^{(2)}_{\rm vector} + S_{\rm GF} 
 &=& \frac{1}{g^2}\,{\rm tr}_{\rm F}
 \int {\rm d}^8 z \,
\Big( 
\vf^\dagger 
\frac{1}{\Box_+} 
(\Box_+ - |\cM_{\rm Ad}|^2 ) \vf 
- \hf 
 v( \Box_{\rm v} - |\cM_{\rm Ad}|^2) v \Big)~.
\label{vector-quad}
\eea

The Feynman propagators 
associated with the actions
(\ref{vector-quad}) and (\ref{hyper-quad}) 
can be expressed via a single Green's function
in  different representations of the gauge group.
Such a Green's function,
$G^{({\rm D})} (z,z')$,  originates 
in the following auxiliary  model
\be
S^{({\rm D})} =
 \int {\rm d}^8 z \,
\S^\dagger
( \Box_{\rm v} - |\cM_{\rm D}|^2) \S~,
\label{S-action}
\ee
which describes the dynamics of an unconstrained complex 
superfield $\S$
transforming in some representation D
of the gauge group. 
The relevant Feynman propagator reads
\be
G^{({\rm D})}(z,z') = 
{\rm i}\,\langle 0| T \, \Big(\S(z)\,  \S^\dagger (z') \Big) |0\rangle
\equiv  {\rm i}\,\langle \S(z) \, \S^\dagger (z')\rangle
\ee
and satisfies the equation
\be
\Big(\Box_{\rm v} - |\cM_{\rm D}|^2  \Big) 
\, G^{({\rm D})}(z,z') 
= - {\bf 1}\,\d^8 (z-z')~.
\label{green}
\ee
The Feynman propagators in the model (\ref{vector-quad})
are
\bea
{ {\rm i} \over g^2} \, \langle  v (z)\, v^{\rm T} (z') \rangle &=& 
-   G^{({\rm Ad})}(z,z') ~, \non \\
{ {\rm i} \over g^2} \, \langle  \vf (z)\,  \vf^\dagger (z') \rangle &=& 
{1 \over 16} {\bar \cD}^2 \cD'^2\, 
G^{({\rm Ad})}(z,z') ~, \\
 \langle  \vf (z)\, \vf^{\rm T} (z') \rangle &=& 
\langle  {\bar \vf} (z)\, \vf^\dagger (z') \rangle
=0~. \non 
\eea
It is understood here that $v$ and $\vf$ 
are column-vectors, and not matrices as in the
preceding consideration.
To formulate the Feynman propagators 
in the model (\ref{hyper-quad}), it is useful 
to introduce the notation 
\bea
{\bf q} = 
\left(
\begin{array}{c}
q\\
\tilde{q}  
\end{array}
\right)~, \qquad 
{\bf q}^\dagger = \Big( q^\dagger, \,\tilde{q}^\dagger \Big)~.  
\eea
Then, the Feynman propagators read
\bea
 {\rm i}  \, \langle  {\bf q}  (z)\,  {\bf q} ^\dagger (z') \rangle &=& 
{1 \over 16} {\bar \cD}^2 \cD'^2\, 
G^{({\rm R} \oplus {\rm R}_{\rm c})}(z,z') ~, \non \\
{\rm i}  \, \langle  q  (z)\,  \tilde{q} ^{\rm T} (z') \rangle &=& 
\cM_{\rm R}^\dagger \,  
G^{({\rm R})}_+(z,z') 
~, \\
{\rm i}  \, \langle  
\overline{\tilde{q}}  (z) \, q ^\dagger (z') \rangle &=&  
\cM_{\rm R} \, 
G^{({\rm R})}_-(z,z') ~,
\non  
\eea
where the covariantly chiral ($G_+$) and antichiral
($G_-$) Green's functions are  related to $G$ as 
follows:
\bea
G_+(z,z') &=& 
-{1 \over 4} {\bar \cD}^2 G(z,z') 
= -{1 \over 4} {\bar \cD}'^2 G(z,z') ~,\non \\
G_-(z,z') &=& 
-{1 \over 4} \cD^2 G(z,z') 
= -{1 \over 4}\cD'^2 G(z,z') ~.
\label{chiral}
\eea

\sect{Functional representation for two-loop supergraphs
with quantum hypermultiplets}

The interactions for the quantum hypermuliplets are: 
\bea
&& S_{\rm int} = \int {\rm d}^8 z \,  v^\m
{\bf q}^\dagger \,\cT_\m \,
{\bf q}
+ \hf  \int {\rm d}^8 z \,  v^\m v^\n
{\bf q}^\dagger \,\cT_\m \cT_\n \,
{\bf q} \non \\
&& \quad - { {\rm i} \over 2} \int {\rm d}^6 z \, \vf^\m \,
{\bf q}^{\rm T} 
\left(
\begin{array}{c c}
0 &  T_\m{}^{\rm T} \\
T_\m& 0  
\end{array}
\right)  
{\bf q} 
+ { {\rm i} \over 2} \int {\rm d}^6 {\bar z} \, 
{\bar \vf}^\m \,
{\bf q}^\dagger
\left(
\begin{array}{c c}
0 &  T_\m{}^{\rm T} \\
T_\m& 0  
\end{array}
\right)  
{\bar {\bf q} }
~,
\eea
where 
\bea 
\cT_\m = \left(
\begin{array}{c c}
T_\m & 0 \\
0& -T_\m{}^{\rm T}  
\end{array}
\right)
\eea
are the generators of the representation 
${\rm R}\oplus {\rm R}_{\rm c}$.

There are four two-loop supergraphs  
with quantum hypermultiplets, 
and they are depicted in Figures 1--4.
\begin{figure}[!htb]
\begin{center}
\includegraphics{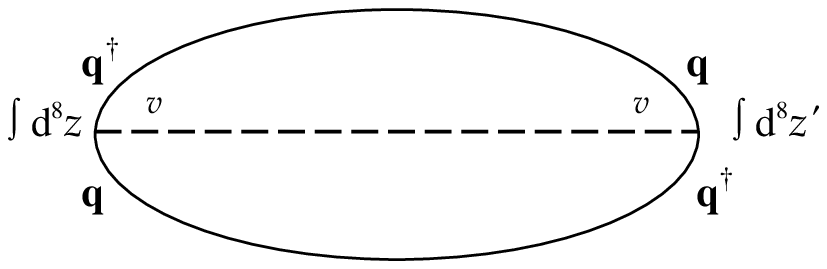}
\caption{Two-loop supergraph I}
\end{center}
\end{figure}
\begin{figure}[!htb]
\begin{center}
\includegraphics{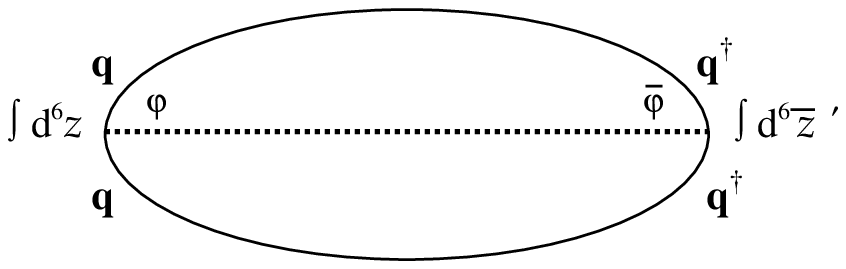}
\caption{Two-loop supergraph II}
\end{center}
\end{figure}

The  contributions from the first two supergraphs 
can be combined in the form 
\bea
\G_{\rm I +II} &=& -{{\rm i} \over 2^9 } \int {\rm d}^8 z 
\int {\rm d}^8 z' \,  \langle  v^\m (z) \,v^\n (z') \rangle  \non \\
&\times &
\, {\rm  tr} \left\{ \cT_\m
\Big( {\bar \cD}^2 \cD^2 \, 
G^{({\rm R} \oplus {\rm R}_{\rm c})}(z,z') \Big) \, \cT_\n
 [{\bar \cD}'^2, \cD'^2 ] \,
G^{({\rm R} \oplus {\rm R}_{\rm c})}(z',z) \right\}~,
\label{I+II}
\eea
where we have used the identities \cite{KM2}
\be
{\bar \cD}^2 \, G(z,z') = {\bar \cD}'^2 \, G(z,z') ~, \qquad 
\cD^2 \, G(z,z') = \cD'^2 \, G(z,z')~.
\label{transpar}
\ee
As in  \cite{KM2}, the above expression can be
considerably simplified.
The representation ${\rm R} \oplus {\rm R}_{\rm c}$ is real, 
\bea
-\cT_\m{}^{\rm T} = \s_1 \, \cT_\m \, \s_1~, 
\qquad 
\s_1= \left(
\begin{array}{c c}
0 \, &\, {\bf 1} \\
{\bf 1} &0  
\end{array}
\right)~.
\label{real-1}
\eea
On the same grounds, the relevant Green's function 
obeys the following reality property:
\be
\Big( G^{({\rm R} \oplus {\rm R}_{\rm c})}(z',z)\Big)^{\rm T}
= \s_1 \, G^{({\rm R} \oplus {\rm R}_{\rm c})}(z,z')\, \s_1~. 
\label{real-2}
\ee
These relations, the symmetry property 
\be
 \langle  v^\m (z)\, v^\n (z') \rangle 
=  \langle  v^\n (z') v^\m (z) \rangle ~,
\ee
and a simple consequence  of (\ref{transpar}), 
\be 
[{\bar \cD}^2, \cD^2 ] \,G(z,z') 
= - [{\bar \cD}'^2, \cD'^2 ] \, G(z,z') ~, 
\ee
allow one to turn (\ref{I+II}) into 
\bea
\G_{\rm I +II} &=& -{{\rm i} \over 2^{10} } \int {\rm d}^8 z 
\int {\rm d}^8 z' \,  \langle  v^\m (z) \,v^\n (z') \rangle  \non \\
&\times &
\, {\rm  tr} \left\{ \cT_\m
\Big([ {\bar \cD}^2, \cD^2] \, 
G^{({\rm R} \oplus {\rm R}_{\rm c})}(z,z') \Big) \, \cT_\n
 [{\bar \cD}'^2, \cD'^2 ] \,
G^{({\rm R} \oplus {\rm R}_{\rm c})}(z',z) \right\}~,
\label{I+II-new}
\eea
Taking into account eqs.  
(\ref{real-1}) and (\ref{real-2}) once again, 
one ends up with 
\bea
\G_{\rm I +II} &=& -{{\rm i} \over 2^{9} } \int {\rm d}^8 z 
\int {\rm d}^8 z' \,  \langle  v^\m (z) \,v^\n (z') \rangle  \non \\
&\times &
\, {\rm  tr} \left\{ T_\m
\Big([ {\bar \cD}^2, \cD^2] \, 
G^{({\rm R})}(z,z') \Big) \, T_\n
 [{\bar \cD}'^2, \cD'^2 ] \,
G^{({\rm R} )}(z',z) \right\}~.
\label{I+II-final}
\eea
The following identity
\be
\frac{1}{16} [\cD^2 , {\bar \cD}^2 ] =   
\frac{\rm i}{4} {\bar \cD}_\ad \cD^{\a \ad } \cD_\a 
-\frac{\rm i}{4} \cD_\a \cD^{\a \ad } {\bar \cD}_\ad~,
\ee
 turns out  to be very useful  when computing the action 
of the commutators of covariant derivatives in 
(\ref{I+II-final}) on  the Green's functions. 

\begin{figure}[!htb]
\begin{center}
\includegraphics{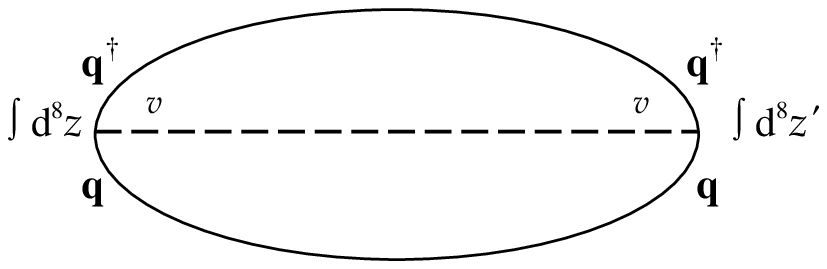}
\caption{Two-loop supergraph III}
\end{center}
\end{figure}

The supergraph in Fig. 3
leads to the following contribution
\bea
\G_{\rm III} &=& {{\rm i} \over 16} \int {\rm d}^8 z 
\int {\rm d}^8 z' \,  \langle  v^\m (z) \, v^\n (z') \rangle  \non \\
&\times &  {\rm  tr} \left\{ T_\m \,\F^\dagger \,
\Big( {\bar \cD}^2  \, G^{({\rm R} )}(z,z') \Big) \,
T_\n \, \F \,  \cD'^2  \,G^{({\rm R} )}(z',z) \right\}~.
\label{III}
\eea

\begin{figure}[!htb]
\begin{center}
\includegraphics{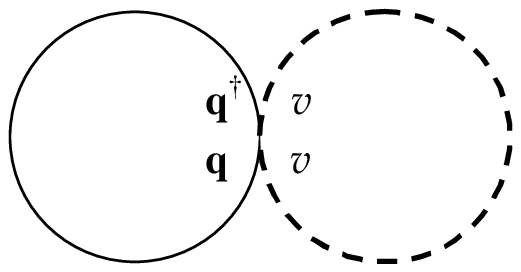}
\caption{Two-loop supergraph IV}
\end{center}
\end{figure}

Finally, the supergraph in Fig. 4
leads to the following contribution
\bea
\G_{\rm IV} &=& -{{\rm i} \over 16} \int {\rm d}^8 z 
\int {\rm d}^8 z' \,  \d^8(z-z')\,
\langle  v^\m (z) \, v^\n (z') \rangle  
\non \\
&\times &  
{\rm  tr} \left\{ T_\m \,T_\n \,
{\bar \cD}^2  \cD'^2 G^{({\rm R} )}(z,z') 
\right\}~.
\label{IV}
\eea

\sect{Exact superpropagators} 
${}$For computing quantum corrections of the form
(\ref{F4}), it is sufficient to consider 
a very special  type of background field configuration
specified by the constraint
\be
\cD_\a \cW_\b = 0~.
\label{glueball}
\ee
This is the simplest representative of 
background vector multiplets for which 
all Feynman superpropagators are known exactly 
 \cite{KM,KM2}. 

${}$For the Green's function 
$G \equiv G^{({\rm R} )}$, 
we introduce the Fock-Schwinger 
proper-time representation 
\be
G(z,z') = 
{\rm i} \int\limits_0^\infty {\rm d}s \, K(z,z'|s) \, 
{\rm e}^{ -{\rm i} (|\cM|^2 -{\rm i}\ve ) s }~, 
\qquad   \ve \to +0~.
\label{proper-time-repr}
\ee
The corresponding heat kernel reads
\be
K(z,z'|s) = -\frac{\rm i}{(4 \pi s)^2} \, 
{\rm e}^{ {\rm i}\r^2 /4s } \, 
\d^2(\z  -{\rm i} s \,\cW) \,
\d^2({\bar \z}  + {\rm  i}s \,{\bar \cW} ) 
\, I(z,z')~,  
\label{real-kernel}
\ee
where the supersymmetric two-point function 
$\z^A(z,z') =- \z^A(z',z)=(\r^a , \z^\a, {\bar \z}_\ad)$
is defined as follows: 
\be
\r^a = (x-x')^a - {\rm i} (\q-\q') \s^a {\bar \q}' 
+ {\rm i} \q' \s^a ( {\bar \q} - {\bar \q}') ~, \quad
\z^\a = (\q - \q')^\a ~, \quad
{\bar \z}_\ad =({\bar \q} -{\bar \q}' )_\ad ~. 
\label{two-point}
\ee

The parallel displacement propagator,
$I(z,z')$, is uniquely specified by 
the following requirements:\\
(i) the gauge transformation law
\be
 I (z, z') ~\to ~
{\rm e}^{{\rm i} \t(z)} \,  I (z, z') \,
{\rm e}^{-{\rm i} \t(z')} ~
\label{super-PDO1}
\ee
with respect to  a  gauge ($\t$-frame) transformation 
of  the covariant derivatives
\be
\cD_A ~\to ~ {\rm e}^{{\rm i} \t(z)} \, \cD_A\,
{\rm e}^{-{\rm i} \t(z)}~, \qquad 
\t^\dagger = \t ~, 
\label{tau}
\ee
with the gauge parameter $\t(z)$ being arbitrary modulo 
the reality condition imposed;\\
(ii) the equation  
\be
\z^A \cD_A \, I(z,z') 
= \z^A \Big( D_A +{\rm i} \, \G_A(z) \Big) I(z,z') =0~;
\label{super-PDO2}
\ee
(iii) the boundary condition 
\be 
I(z,z) ={\bf 1}~.
\label{super-PDO3}
\ee
These imply the important relation
\be
I(z,z') \, I(z', z) = {\bf 1}~,
\label{collapse}
\ee
as well as 
\be
\z^A \cD'_A \, I(z,z') 
= \z^A  \Big( D'_A \,I(z,z') 
 - {\rm i} \, I(z,z') \, \G_A(z') \Big) =0~.
\ee

${}$For the background (\ref{glueball}), 
the parallel displacement propagator is completely 
specified by the properties:
\bea
I(z',z) \, \cD_{\a \ad} I(z,z') &=&
-{\rm i} ( \z_{\a} \bar{\cW}_{\ad} 
+ \cW_\a \, \bar{\z}_{\ad}   ) ~, \non \\
I(z',z) \,  \cD_{\a} I(z,z') &=&
- \frac{{\rm i}}{2} \, \r_{\a \ad} 
\bar{\cW}^{\ad} 
+ \frac13 ( \z_{\a} \bar{\z} \bar{\cW}
+ \bar{\z}^2  \cW_{\a})~, \\
I(z',z) \,  {\bar \cD}_{\ad} I(z,z') &=&
- \frac{{\rm i}}{2} \, \r_{\a \ad} \cW^{\a} 
- \frac13 ( {\bar \z}_\ad \z \cW  + \z^2 {\bar \cW}_\ad)~. \non
\eea

The heat kernel corresponding 
to the chiral Green's function 
$G_+$ (\ref{chiral}) is
\bea
K_+(z,z'|s) &=&-{1 \over 4} {\bar \cD}^2 K(z,z'|s) \non \\
 &=& -\frac{\rm i}{(4 \pi s)^2} \, 
{\rm e}^{ {\rm i}\r^2 /4s } \, 
\d^2(\z  -{\rm i} s \,\cW) \,
{\rm e}^{ \frac{{\rm i}}{6} s\,\cW^2 \,({\bar \z} +{\rm i}s\, {\bar \cW})^2}
\, I(z,z')~. 
\label{chiral-glueball}
\eea
It is an instructive exercise to check, using the properties
of the parallel displacement propagator, 
that $K_+(z,z'|s) $ is covariantly chiral in both arguments. 

The supersymmetric theories that we are going 
to study below  are free of ultraviolet divergences. 
This does not mean that individual 
(say, two-loop) supergraphs are all finite; only their sum,  
at any loop order, has 
to be finite. To deal with UV divergent supergraphs, 
we will adopt  supersymmetric dimensional regularization 
via dimensional reduction \cite{GGRS}. 
All manipulations with the gauge covariant derivatives 
(D-algebra) have to be completed in four dimensions. 
At a final stage, the bosonic part of the heat kernel 
(\ref{real-kernel}) is to be continued to $d$ dimensions
using the prescription
\bea
\frac{\rm i}{(4 \pi {\rm i} s)^2} \, 
{\rm e}^{ {\rm i}\r^2 /4s }  \quad \longrightarrow \quad 
\frac{\rm i}{(4 \pi {\rm i} s)^{d/2}} \, 
{\rm e}^{ {\rm i}\r^2 /4s } ~,  \qquad \quad 
d= 4-\ve~.
\eea
It is assumed that loop space-time integrals
are done in $d$ dimensions, 
using the following integration rules:
\bea
\frac{\rm i}{(4 \pi {\rm i} )^{d/2}} 
\int {\rm d}^d \r \, {\rm e}^{ {\rm i} C \r^2 /4 } 
&=&  C^{-d/2}~, 
\non  \\
 \frac{\rm i}{(4 \pi {\rm i} )^{d/2}} 
\int {\rm d}^d \r \, \r_a \r_b\, 
{\rm e}^{ {\rm i} C \r^2 /4 } 
&=& 2\,{\rm i} \, \eta_{ab}\,
 C^{-(d/2+1)}~,
\label {int-rules}
\eea
with $C$ a positive parameter. 

\sect{$SU(N)$ SYM with $2N$ hypermultiplets
in the fundamental}
${}$From now on, we choose the gauge group to be $SU(N)$.
Lower-case Latin letters from the middle of the alphabet, 
$i,j,\dots$, 
will be used to denote matrix elements in the fundamental, 
with the convention $i =0,1,\dots, N-1 \equiv 0, \un{ i}$.
We choose a Cartan-Weyl basis 
to consist of the elements:
\be 
H_I = \{ H_0, H_{\un{I}}\}~, \quad  \un{I} = 1,\dots, N-2~, 
\qquad \quad E_{ij}~, \quad i\neq j~. 
\label{C-W}
\ee 
The basis elements in the fundamental representation 
are defined similarly to \cite{Georgi}, 
\bea
(E_{ij})_{kl} &=& \d_{ik}\, \d_{jl}~, \non \\
(H_I)_{kl} &=& \frac{1}{\sqrt{(N-I)(N-I-1)} }
\Big\{ (N-I)\, \d_{kI} \, \d_{lI} - 
\sum\limits_{i=I}^{N-1} \d_{ki} \, \d_{li} \Big\} ~,
\eea
and are characterized by the properties
\be
{\rm tr}_{\rm F} (H_I\,H_J) = \d_{IJ}~, 
\qquad 
{\rm tr}_{\rm F} (E_{ij}\,E_{kl}) = \d_{il}\,\d_{jk}~, 
\qquad {\rm tr}_{\rm F} (H_I \,E_{kl}) =0~.
\ee

The $\cN=2$ background vector multiplet is chosen to be
\be
\F = \f \, H_0~, \qquad \cW_\a = W_\a \, H_0~,
\ee
Its characteristic feature is that it leaves
the subgroup $U(1) \times SU(N-1) \subset SU(N)$ 
unbroken, where $U(1)$ is associated with $H_0$
and  $SU(N-1)$ is  generated by 
$\{ H_{\un{I}}, \, E_{ \un{i} \un{j}} \}$.
In evaluating the supergraphs, we 
consider $\f$ and $W_\a$ to be constant.
This suffices for our purposes.

The mass matrix is 
\be
|\cM|^2 = {\bar \f}\f \, (H_0)^2~, 
\ee
and therefore a superfield's mass is determined 
by  its $U(1)$ charge with respect to $H_0$. 
With the notation 
\be
e_{\rm f} = \sqrt{ { N-1 \over N}}  ={1 \over e_{\rm a} }~,
\ee
the $U(1)$ charges of all quantum superfields 
are given in the table.
\begin{center}
\begin{tabular}{ | c || c | c || c | c| |c| c| c|}  \hline
superfield & $q_0$ & $q_{\un{i}}$ & $\tilde{q}_0 $ &
$\tilde{q}_{\un{i}} $ & $v^{0\, \un{i} }$ & $v^I$ &
$v^{ \un{i} \,\un{j} }$\\ \hline 
$U(1) $ charge & $e_{\rm f} $ & $e_{\rm f} - e_{\rm a} $ &
$-e_{\rm f} $ & $e_{\rm a} - e_{\rm f} $ & $e_{\rm a} $ 
&0&0\\  \hline
\end{tabular} \\
${}$\\
Table  1: $U(1)$ charges of superfields
\end{center}
As can be  seen, all fundamental hypermultiplet superfields are 
massive. For the adjoint superfields\footnote{Since the basis 
(\ref{C-W}) is not orthonormal, 
${\rm tr}_{\rm F} (T_\m \, T_\n)  =g_{\m \n} \neq \d_{\m \n}$, 
it is necessary to keep track of the Cartan-Killing metric when 
working with adjoint vectors. For any elements
$u=u^\m T_\m $ and $v=v^\m T_\m $ of the Lie algebra, 
we have $u\cdot v =  {\rm tr}_{\rm F} (u\,v) =u^\m \,v_\m$, 
where $v_\m = g_{\m\n} v^\n$
($v_I =v^I$, $v_{ij}= v^{ji}$).}
\be
v = v^I \, H_I + v^{ij} \, E_{ij} \equiv v^\m \,T_\m~,  
\qquad i \neq j ~, 
\ee
there are $2(N-1)$ massive superfields 
($v^{0 \un{i}}$ and their conjugates $v^{ \un{i}0}$), 
while the remaining  $(N-1)^2$ superfields, $v^I$ and 
$v^{\un{i} \,\un{j}} $,  are free massless.
This follows from the identity 
\be
[H_0 , E_{ij}] ~=~ \sqrt{N \over N-1}\,  \Big( 
\d_{0i}\, E_{0j} - \d_{0j}\, E_{i0} \Big)~. 
\ee

Let us denote by $\bG^{(e)}(z,z')$ the Green's function 
(\ref{green}) in the special  case when 
the gauge group is $U(1)$ generated by $H_0$, 
and the quantum superfield $\S$ in (\ref{S-action})
carries $U(1)$ charge $e$, 
$H_0 \,\S = e\, \S$ (in particular, the mass matrix is 
$|\cM|^2 = e^2 \, {\bar \f} \f$). 
The Green's function has the proper-time representation
\be
\bG^{(e)}(z,z') = 
{\rm i} \int\limits_0^\infty {\rm d}s \, 
\bK^{(e)}(z,z'|s) \, 
{\rm e}^{ -{\rm i} ( e^2  {\bar \f} \f  -{\rm i}\ve ) s }~, 
\qquad   \ve \to +0~,
\label{U(1)GF}
\ee
where the  heat kernel is 
\be
\bK^{(e)}(z,z'|s) = -\frac{\rm i}{(4 \pi s)^2} \, 
{\rm e}^{ {\rm i}\r^2 /4s } \, 
\d^2(\z  -{\rm i} es \,W) \,
\d^2({\bar \z}  + {\rm  i}es \,{\bar W} ) 
\, I^{(e)}(z,z')~.
\label{U(1)HK}
\ee
${}$For the $\cN=2$ background vector multiplet chosen, 
all the Feynman propagators are expressed via such 
$U(1)$ Green's functions. 

In the remainder of this section, we 
specialize to  the case of $\cN=2$ SYM
with $2N$ hypermultiplets in the fundamental 
representation of $SU(N)$. This theory is finite
since the finiteness condition 
(\ref{finita}) is satisfied 
due to the well-known $SU(N)$ identity
\be
{\rm tr}_{\rm Ad} \,\J^2 ~=~ 2N\, {\rm tr}_{\rm F} \, \J^2~, 
\qquad  \quad \J \in sl(N)~. 
\ee

\subsection{Evaluation of $\G_{\rm I+II}$ }
We now turn to evaluating $\G_{\rm I+II}$. 
In accordance with (\ref{I+II-final}),
it is necessary to analyze the expression 
\bea
2N\,  \langle  v^\m (z) \,v^\n (z') \rangle  
 &{\rm  tr}_{\rm F} & \left\{ T_\m
\Big([ {\bar \cD}^2, \cD^2] \, 
G^{({\rm F})}(z,z') \Big) \, T_\n
 [{\bar \cD}'^2, \cD'^2 ] \,
G^{({\rm F} )}(z',z) \right\} \non \\
= 2N \sum\limits_{I} 
\langle  v^I (z) \,v^I (z') \rangle  
& {\rm  tr}_{\rm F} & \left\{ H_I
\Big([ {\bar \cD}^2, \cD^2] \, 
G^{({\rm F})}(z,z') \Big) \, H_I 
 [{\bar \cD}'^2, \cD'^2 ] \,
G^{({\rm F} )}(z',z) \right\} 
\label{I+II-fund1}\\
+ 2N\sum\limits_{i\neq j} 
\langle  v^{ij}  (z) \,v^{ji} (z') \rangle  
&{\rm  tr}_{\rm F} &\left\{ E_{ij}
\Big([ {\bar \cD}^2, \cD^2] \, 
G^{({\rm F})}(z,z') \Big) \, E_{ji} 
 [{\bar \cD}'^2, \cD'^2 ] \,
G^{({\rm F} )}(z',z) \right\} ~, \non 
\eea
where the factor $2N$ relates to the presence 
of $2N$ hypermultiplets.
The expression in the second line can be 
simplified on the basis of 
the following observations: (i) the propagator 
$G^{({\rm F} )}$ is diagonal; (ii)
the massless adjoint  propagators are identical, 
\be 
\langle  v^I (z) \,v^I (z') \rangle  = 
\langle  v^{\un{i}\, \un{j}}  (z) \,v^{\un{j}\,\un{i}} (z') \rangle  
= {\rm i}\,g^2 \,{\bf G}^{(0)} (z,z')~.
\ee
with ${\bf G}^{(0)} (z,z')$ the free massless Green's function.
Then, the second line of (\ref{I+II-fund1}) becomes
\be 
2{\rm i}N \,g^2\,\bG^{(0)} (z, z')  \, 
\sum\limits_{I}  
 {\rm  tr}_{\rm F}  \left\{ (H_I)^2
\Big([ {\bar \cD}^2, \cD^2] \, 
G^{({\rm F})}(z,z') \Big) \,
 [{\bar \cD}'^2, \cD'^2 ] \,
G^{({\rm F} )}(z',z) \right\}~. 
\ee
In the fundamental representation of $SU(N)$,  
\be 
\sum\limits_{I} (H^{({\rm F})}_I)^2 ~=~ { {N-1} \over N} \, {\bf 1}~.
\ee 
This gives
\bea
&& 2N \sum\limits_{I} 
\langle  v^I (z) \,v^I (z') \rangle  
 {\rm  tr}_{\rm F}  \left\{ H_I
\Big([ {\bar \cD}^2, \cD^2] \, 
G^{({\rm F})}(z,z') \Big) \, H_I 
 [{\bar \cD}'^2, \cD'^2 ] \,
G^{({\rm F} )}(z',z) \right\} \non \\
&=& 2 {\rm i}\,( N-1) \, g^2\,
\bG^{(0)} (z,z')  \, 
\Big\{ \Big([ {\bar \cD}^2, \cD^2] \, 
\bG^{(e_{\rm f})}(z,z') \Big) \,
 [{\bar \cD}'^2, \cD'^2 ] \,
\bG^{(e_{\rm f} )}(z',z) \\
&&\qquad \qquad +~ (N-1) 
\Big([ {\bar \cD}^2, \cD^2] \, 
\bG^{(e_{\rm f}-e_{\rm a} )}(z,z') \Big) \,
 [{\bar \cD}'^2, \cD'^2 ] \,
\bG^{(e_{\rm f} -e_{\rm a})}(z',z) \Big\} ~. \non 
\eea

To transform the expression in the 
third  line of  (\ref{I+II-fund1}),  
we notice 
\bea
&& {\rm  tr}_{\rm F} \left\{ E_{ij} \Big([ {\bar \cD}^2, \cD^2] \, 
G^{({\rm F})}(z,z') \Big) \, E_{ji}  [{\bar \cD}'^2, \cD'^2 ] \,
G^{({\rm F} )}(z',z) \right\} \non \\
&&\quad = \Big([ {\bar \cD}^2, \cD^2] \, 
G^{({\rm F})}(z,z') \Big)_{jj} \, 
\Big(  [{\bar \cD}'^2, \cD'^2 ] \,
G^{({\rm F} )}(z',z) \Big)_{ii}~, 
\eea
as immediately follows from the definition of $E_{ij}$.
This leads to 
\bea
&& \sum\limits_{i\neq j} 
\langle  v^{ij}  (z) \,v^{ji} (z') \rangle  
{\rm  tr}_{\rm F} \left\{ E_{ij}
\Big([ {\bar \cD}^2, \cD^2] \, 
G^{({\rm F})}(z,z') \Big) \, E_{ji} 
 [{\bar \cD}'^2, \cD'^2 ] \,
G^{({\rm F} )}(z',z) \right\}   \\
&=& {\rm i} \, (N-1)g^2\,  \bG^{(e_{\rm a} )}(z,z') \,
\Big([ {\bar \cD}^2, \cD^2] \, 
\bG^{(e_{\rm f} -e_{\rm a})}(z,z') \Big) \,
[{\bar \cD}'^2, \cD'^2 ] \,
\bG^{(e_{\rm f} )}(z',z)  ~+~
 (z \leftrightarrow z') \non \\
&&+ {\rm i}\, (N-1)(N-2) g^2\,
\bG^{(0)} (z, z') \, 
\Big([ {\bar \cD}^2, \cD^2] \, 
\bG^{(e_{\rm f} -e_{\rm a})}(z,z') \Big) \,
[{\bar \cD}'^2, \cD'^2 ] \,
\bG^{(e_{\rm f} - e_{\rm a} )}(z',z) ~. \non 
\eea

As should be clear from the above consideration,
the evaluation of $\G_{\rm I+II} $ amounts to computing 
a functional of the form 
\be
\int {\rm d}^8 z  \int {\rm d}^8 z' \,  
\bG^{(e_1)} (z, z') \, 
\Big([ {\bar \cD}^2, \cD^2] \, 
\bG^{(e_2)}(z,z') \Big) \,
[{\bar \cD}'^2, \cD'^2 ] \,
\bG^{(e_1 + e_2 )}(z',z) ~, 
\label{block}
\ee 
for some charges $e_1$ and $e_2$. 
${}$For all the Green's functions, we introduce
the proper-time representation (\ref{U(1)GF}). 
Due to the explicit structure of the heat 
kernel, eq. (\ref{U(1)HK}), the first 
multiplier in (\ref{block}) contains a 
Grassmann delta-function,  
$$
\d^2(\z  -{\rm i} e_1s \,W) \,
\d^2({\bar \z}  + {\rm  i}e_1s \,{\bar W} ) ~, 
$$
which allows us to do the integral over
$\q'$.  Next, the second and third multipliers in 
(\ref{block}) can be evaluated 
(in $d$ dimensions) as follows:
\bea
\frac{1}{16} [ {\bar \cD}^2, \cD^2 ] \, 
\bK^{(e)}(z,z' | s) 
& \approx  & \phantom{-}
\frac{\rm i}{(4 \pi s)^2} \, 
{\r  ^{\a \ad} \over s}
(\z -{\rm i} es W)_\a   
 (\bar{\z} +{\rm i} es {\bar W})_\ad  \,
{\rm e}^{ {\rm i} \r^2/4s} \, I^{(e)}(z,z')  \non \\
\longrightarrow
& & - \frac{\rm i}{(4 \pi {\rm i}s)^{d/2}} \, 
{\r  ^{\a \ad} \over s}
(\z -{\rm i} es W)_\a   
 (\bar{\z} +{\rm i} es {\bar W})_\ad  \,
{\rm e}^{ {\rm i} \r^2/4s} \, I^{(e)}(z,z') ~,\non 
\eea
where we have omitted all terms of at least third order 
in the Grassmann variables  $\z_\a, \,{\bar \z}_\ad$ and 
$W_\a, \, {\bar W}_\ad$ 
as they  do not contribute to (\ref{block}). 
Now, the parallel displacement propagators 
associated with the three Green's functions
in (\ref{block}) simply annihilate each other. 
Finally , the integral over $x'$ in (\ref{block}) 
can easily be done if one  first replaces the bosonic variables
$\{x,x'\} \to \{x,\r\}$ and then applies eq. 
(\ref{int-rules}). 
Of special importance is the fact that 
the functional 
\be
\J = {1\over 2^8}
\int {\rm d}^8 z  \int {\rm d}^8 z' \,  
\bG^{(0)} (z, z') \, 
\Big([ {\bar \cD}^2, \cD^2] \, 
\bG^{(e)}(z,z') \Big) \,
[{\bar \cD}'^2, \cD'^2 ] \,
\bG^{(e )}(z',z) 
\label{FUN-1}
\ee 
is finite (so we set $d = 4$) and does not
depend on the charge $e$,
for the background field configuration chosen,
\bea
\J &= & {4e^4 \over (4\p)^4 }
\int {\rm d}^8 z  \, W^2 {\bar W}^2 \, 
\int\limits_{0}^{\infty} 
{\rm d}s_1  
\int\limits_{0}^{\infty}
 {\rm d}s_2 
\int\limits_{0}^{\infty} 
{\rm d}s_3  \,
{\rm e}^{-e^2 \f {\bar \f} (s_2+s_3)} \,
\frac{s_1\,s_2{}^2 \,s_3{}^2} 
{(s_2s_3 + s_1s_2 +s_1s_3)^3} \non \\
&=& {1 \over 3}\, {1 \over (4\p)^4 }
\int {\rm d}^8 z  \, 
{ W^2 {\bar W}^2  \over (\f {\bar \f})^2}
=   {1 \over 3}\, {1 \over (4\p)^4 } \, \U~,
\label{FUN-2}
\eea
see \cite{KM2} for more details.

The computational scheme outlined 
leads to the final result\footnote{In this paper, 
we do not evaluate all of the proper-time 
integrals,  such as $I_{\rm I+II} $. 
We are only interested in their large $N$ behaviour 
and in their singular parts. This is why we freely 
set $d=4$ in  finite multiplicative factors, such as
$(\f {\bar \f})^{-d/2}$. No mass scale 
is required because the total contribution is finite. }: 
\be
\G_{\rm I+II} = \frac{g^2}{(4\p)^4} \, 
\U \Big\{ {1\over 3} \,N(N-1)^2  
+  8\,N(N-1)\, I_{\rm I+II} \Big\} ~,
\ee
where 
\bea
 I_{\rm I+II}  = \int\limits_{0}^{\infty} {\rm d}s_1 
{\rm d}s_2
{\rm d}s_3 \, 
{\rm e}^{-[{\rm e}_{\rm a}^2 s_1 
+ ({\rm e}_{\rm a}-{\rm e}_{\rm f})^2 s_3
+ {\rm e}_{\rm f}^2 s_2]} \, 
\frac{ s_1 
[{\rm e}_{\rm a} s_1 + ({\rm e}_{\rm a}-{\rm e}_{\rm f})s_3]^2
[{\rm e}_{\rm a} s_1 + {\rm e}_{\rm f}s_2 ]^2}
{[s_2s_3 +s_1s_2 +s_1 s_3]^{d/2 + 1} }
\label{I-I+II}
\eea
is a divergent integral
in the limit $\ve = 4-d \to 0$. 

To isolate the divergence in (\ref{I-I+II}),
we first rescale the  integral 
\bea
 I_{\rm I+II}  &=& {(N-1)^4 \over N^2} 
\int\limits_{0}^{\infty} {\rm d}t_1  
{\rm d}t_2
{\rm d}t_3 \, 
{\rm e}^{-[ t_1 + t_2 + t_3]} \,
\frac{ t_1 
[t_1 + Nt_3]^2
[t_1 + {N \over N-1} t_2 ]^2}
{[t_1 t_2 + (N-1)^2t_1t_3 + N^2 t_2 t_3]^{d/2 + 1} }~.
\eea
Now,  it is advantageous to  
introduce new variables \cite{Zinn}:
\be
t_1 = s\,t\,u~, \qquad 
t_2 =s\,t\,(1-u)~, \qquad 
t_3 =s\,(1-t)~, 
\ee
with the important properties $s = t_1 +t_2+t_3$, 
$s\,t = t_1 +t_2$ and 
\be 
\int\limits_{0}^{\infty} {\rm d}t_1 
\int\limits_{0}^{\infty} {\rm d}t_2 
\int\limits_{0}^{\infty} {\rm d}t_3 \cdots 
~=~\int\limits_{0}^{\infty} {\rm d}s
\int\limits_{0}^{1} {\rm d} t 
\int\limits_{0}^{1} {\rm d} u \, s^2 \,t  \cdots
\ee
The integral over $s$ factorizes and it is convergent, 
\be
\int\limits_{0}^{\infty} {\rm d}s \,
s^{5-d}\, {\rm e}^{-s} = 1 +O(\ve)~.
\ee
As a result , we obtain
\bea
I_{\rm I+II}  &=& \left({N-1 \over N} \right)^2
\int\limits_{0}^{1} {\rm d} t 
\int\limits_{0}^{1} {\rm d} u
\frac{ t^{3-d/2} u [N-u]^2 [N-t(N-u)]^2}
{[(N-u)^2(1-t) +u(1-u) ]^{d/2 +1} } ~+~{\rm finite}~.
\eea
The divergent part of $I_{\rm I+II}$
turns out to be 
\be
(I_{\rm I+II})_{\rm div} = 
\left(-1 +4 \, {N-1 \over N^2} \right)\,
 {1 \over \ve} ~.
\label{I-I+II-div}
\ee

\subsection{Evaluation of $\G_{\rm III}$ }

The evaluation of $\G_{\rm III}$ is very similar 
to that of $\G_{\rm I+II}$ just described. 
Therefore, we simply give the final result: 
\be
\G_{\rm III} = \frac{g^2}{(4\p)^4} \, 
\U \Big\{ {2\over 3} \,N(N-1)^2  
-  4\,(N-1)\, I_{\rm III} \Big\} ~,
\ee
where 
\bea
 I_{\rm III}  = \int\limits_{0}^{\infty} {\rm d}s_1 
{\rm d}s_2
{\rm d}s_3 \, 
{\rm e}^{-[{\rm e}_{\rm a}^2 s_1 
+ ({\rm e}_{\rm a}-{\rm e}_{\rm f})^2 s_3
+ {\rm e}_{\rm f}^2 s_2]} \, 
\frac{  
[{\rm e}_{\rm a} s_1 + ({\rm e}_{\rm a}-{\rm e}_{\rm f})s_3]^2
[{\rm e}_{\rm a} s_1 + {\rm e}_{\rm f}s_2 ]^2}
{[s_2s_3 +s_1s_2 +s_1 s_3]^{d/2 } }
\label{I-III}
\eea
is a divergent integral
in the limit $\ve = 4-d \to 0$. 
Its divergent part proves to be
\be
(I_{\rm III})_{\rm div} =  
4 \, {N-1 \over N} \,
 {1 \over \ve} ~.
\ee

\subsection{Evaluation of $\G_{\rm IV}$ }
It remains to evaluate  $\G_{\rm IV}$. 
As is seen from  its defining  expression (\ref{IV}), 
$ \G_{\rm IV}$ involves a vector propagators 
at coincident points,  
$\langle  v^\m (z) \, v^\n (z) \rangle $. 
The latter is non-trivial only for the 
massive  superfields, 
\bea 
\langle  v^I (z) \, v^I (z) \rangle &=&
\langle  v^{\un{i} \un{j} }  (z) \, 
v^{\un{j} \un{i} } (z) \rangle =0~, \non \\
\langle  v^{0 \un{i}  }  (z) \, 
v^{\un{i}0 } (z) \rangle &=& 
\langle  v^{\un{i} 0 }  (z) \, 
v^{0  \un{i} } (z) \rangle
=-{ g^2 \over 8 \p^2  } 
\left( {N-1 \over N} \right) \,
\frac{ W^2 {\bar W}^2 }{ (\f {\bar \f})^3 } ~. 
\eea
Thus, we can rewrite $ \G_{\rm IV}$ in the form 
\bea
\G_{\rm IV} &=& {{\rm i}g^2  \over 128 \p^2 }
\left({N-1 \over N} \right) 
 \int {\rm d}^8 z 
\frac{ W^2 {\bar W}^2 }{ (\f {\bar \f})^3 } \,
{\rm  tr} \left\{ \{ E_{0 \un{i} }, E_{\un{i} 0} \} \,
{\bar \cD}^2  \cD'^2 G^{({\rm R} )}(z,z') \Big|_{z'=z}
\right\}~.
\label{IV-mod}
\eea

In the superconformal theory under consideration, 
the quantum correction  (\ref{IV-mod}) is
\bea
\G_{\rm IV} &=& {{\rm i}g^2  \over 64 \p^2 }
(N-1 ) 
 \int {\rm d}^8 z 
\frac{ W^2 {\bar W}^2 }{ (\f {\bar \f})^3 } \,
{\rm  tr}_{\rm F} 
\left\{ \{ E_{0 \un{i} }, E_{\un{i} 0} \} \,
{\bar \cD}^2  \cD'^2 G^{({\rm F} )}(z,z') \Big|_{z'=z}
\right\}~.
\eea
Its direct evaluation gives
\be
\G_{\rm IV} = -4
\frac{g^2}{(4\p)^4} \, 
\U \,(N-1) \Big\{ N -  2\,{N-1 \over N} \Big\}
\, I_{\rm IV}  ~,
\ee
where 
\be
 I_{\rm IV}  = \int\limits_{0}^{\infty} {\rm d}s
 \,s^{-d/2} \,  {\rm e}^{-s} 
\label{IV-int}
\ee
is a a divergent integral
in the limit $\ve = 4-d \to 0$, 
\be
(I_{\rm IV})_{\rm div} =  
- {2 \over \ve} ~. 
\ee
It is easy to check that 
\be
\Big( \G_{\rm I+II} +
\G_{\rm III} +
\G_{\rm IV}  \Big)_{\rm div} = 0 ~,
\ee
consistent with the finiteness of the theory.

\sect{$\cN=4$ SYM}

We now turn to evaluating to the two-loop supergraphs
with quantum hypermultiplets in the $\cN=4$ super Yang-Mills
theory which is simply $\cN=2$ SYM with a single hypermultiplet 
in the adjoint.

\subsection{Evaluation of $\G_{\rm I+II}$ }
We start by analyzing 
$\G_{\rm I+II}$ 
in the case of the adjoint representation. 
According to (\ref{I+II-final}), we have to compute  
\bea
&&  \langle  v^\m \,v'^\n \rangle  
 {\rm  tr}_{\rm Ad}  \left\{ T_\m
\hat{G} \, T_\n \, \hat{G}'  \right\} 
\label{I+II-ad1}
\eea
where we have introduced the following 
condensed  notation:
$$
 \langle  v^\m \,v'^\n \rangle  
= \langle  v^\m (z) \,v^\n (z') \rangle ~, \quad 
\hat{G} =  [ {\bar \cD}^2, \cD^2] \,  G^{({\rm Ad})}(z,z')~,
\quad 
\hat{G}' = [ {\bar \cD}'^2, \cD'^2] \,  G^{({\rm Ad})}(z',z)~.
$$
Relative to the basis 
$T_\m = (H_I, E_{0 \un{i} }, E_{\un{i} 0} , E_{ \un{i} \un{j} })$, 
the hypermultiplet operator 
$\hat{G} = (\hat{G}^\l{}_\r) $ in (\ref{I+II-ad1})
has a diagonal structure, 
\be
\hat{G} = {\rm diag} \Big(\hat{\bG}^{(0)}\, {\bf 1}_{N-1} , ~
\hat{\bG}^{(e_{\rm a} )}\, {\bf 1}_{N-1}, ~
\hat{\bG}^{(-e_{\rm a} )}\, {\bf 1}_{N-1}, ~
\hat{\bG}^{(0)}\, {\bf 1}_{(N-1)(N-2)} \Big)~,
\label{I+II-ad12}
\ee
with the $U(1)$ charges given explicitly. 
The evaluation of (\ref{I+II-ad1}) will be based on 
considerations of charge conservation. 
At each vertex ($z$ or $z'$), the total charge must be zero. 
The possible charges in the adjoint representation are:
$0, \pm e_{\rm a}$. Therefore, there are 
 contributions to (\ref{I+II-ad1})
of the two different types: (i) one line is neutral,
and hence free massless, while the other two lines carry charges 
$\pm   e_{\rm a}$; (ii) all three lines are neutral, and hence 
free massless. The case (ii) can safely be ignored since 
no dependence on the background fields is present.
With such considerations in mind, we first separate
the contributions to (\ref{I+II-ad1}) with neutral
and charged gauge field lines:
\bea
&&  \sum\limits_{I} \langle  v^I \,v'^I \rangle  
 {\rm  tr}_{\rm Ad}  \left\{ H_I
\hat{G} \,   H_I \,\hat{G}'  \right\} 
+ \sum\limits_{\un{i} \neq \un{j}} 
\langle  v^{\un{i}\un{j} }  \,v'^{\un{j} \un{i} }  \rangle  
 {\rm  tr}_{\rm Ad}  \left\{ E_{\un{i} \un{j} }
\hat{G} \,   E_{ \un{j} \un{i} }  \,\hat{G}'  \right\} \non \\
&+&   
\sum\limits_{\un{i} } 
\langle  v^{0 \un{i} }  \,v'^{ \un{i} 0  }  \rangle  
 {\rm  tr}_{\rm Ad}  \left\{ E_{0 \un{i}  }
\hat{G} \,  E_{ \un{i} 0 }  \,\hat{G}'  \right\}  
+ \sum\limits_{\un{i} } 
\langle  v^{ \un{i}0 }  \,v'^{ 0\un{i}  }  \rangle  
 {\rm  tr}_{\rm Ad}  \left\{ E_{ \un{i} 0 }
\hat{G} \,  E_{ 0\un{i}  }  \,\hat{G}'  \right\}  ~.
\label{I+II-ad3}
\eea
Since the propagators 
$\langle  v^I \,v'^I \rangle  =
\langle  v^{\un{i}\un{j} }  \,v'^{\un{j} \un{i} }  \rangle 
={\rm i}\,g^2 {\bf G}^{(0)} (z,z')$
are free massless, both $\hat{G}$ and $\hat{G}'$ 
in the first line of (\ref{I+II-ad3}) should be charged. 
In the second line of (\ref{I+II-ad3}), one of the
$\hat{G}$ and $\hat{G}'$ should be  neutral,
while  the other is charged. 
We will analyze separately 
the contributions  appearing in (\ref{I+II-ad3}).

Let $T^{({\rm Ad})}_\m$ be the matrix generators 
in the adjoint representation, 
\be
[T_\m ,  T_\n ] = T_\l \, 
(T^{({\rm Ad})}_\m )^\l{}_\n ~.
\ee
Since $H_I$, $\hat{G}$ and $\hat{G}'$ are diagonal, 
the first term in (\ref{I+II-ad3}) becomes 
\bea
&& \sum\limits_{I} \langle  v^I \,v'^I \rangle  
 {\rm  tr}_{\rm Ad}  \left\{ H_I
\hat{G} \,   H_I \,\hat{G}'  \right\} 
= \sum\limits_{I} \langle  v^I \,v'^I \rangle  
 {\rm  tr}_{\rm Ad}  \left\{ (H_I)^2
\hat{G} \,  \hat{G}'  \right\} \non \\
&=& {\rm i} \,g^2\,{\bf G}^{(0)} (z,z') \Big\{ 
\hat{\bG}^{(e_{\rm a} )}\,\hat{\bG}'^{(e_{\rm a} )}
+ \hat{\bG}^{(-e_{\rm a} )}\,\hat{\bG}'^{(-e_{\rm a} )}\Big\}
\sum\limits_{I} \sum\limits_{\un{i} }  
(H^{({\rm Ad})}_I)^{0\un{i}} {}_ {0\un{i}} \,
(H^{({\rm Ad})}_I)^{0\un{i}} {}_ {0\un{i}} ~, 
\eea
where we have used the identity
\be
(H^{({\rm Ad})}_I)^{0\un{i}} {}_ {0\un{i}} 
=-(H^{({\rm Ad})}_I)^{\un{i}0} {}_ {\un{i}0} ~.
\ee
The group-theoretic factor in the last expression 
is easy to evaluate:
\be
\sum\limits_{I} \sum\limits_{\un{i} } 
(H^{({\rm Ad})}_I)^{0\un{i}} {}_ {0\un{i}} \,
(H^{({\rm Ad})}_I)^{0\un{i}} {}_ {0\un{i}}  = 2(N-1) ~.
\label{GT-1}
\ee

Let us turn to the second term in (\ref{I+II-ad3}), 
\be
\sum\limits_{\un{i} \neq \un{j}} 
\langle  v^{\un{i}\un{j} }  \,v'^{\un{j} \un{i} }  \rangle  
 {\rm  tr}_{\rm Ad}  \left\{ E_{\un{i} \un{j} }
\hat{G} \,   E_{ \un{j} \un{i} }  \,\hat{G}'  \right\} 
={\rm i} \, g^2\, {\bf G}^{(0)} (z,z') \,
\sum\limits_{\un{i} \neq \un{j}} 
 {\rm  tr}_{\rm Ad}  \left\{ E_{\un{i} \un{j} }
\hat{G} \,   E_{ \un{j} \un{i} }  \,\hat{G}'  \right\} ~.
\ee
Since both the hypermultiplets must be 
massive and of opposite charge, 
for this expression we get 
\bea
{\rm i} \,g^2\, {\bf G}^{(0)} (z,z') \Big\{ 
\hat{\bG}^{(e_{\rm a} )}\,\hat{\bG}'^{(e_{\rm a} )}
+ \hat{\bG}^{(-e_{\rm a} )}\,\hat{\bG}'^{(-e_{\rm a} )}\Big\}
 \sum\limits_{\un{i} \neq \un{j}  }
 \sum\limits_{\un{k} ,\, \un{l}  } 
(E^{({\rm Ad})}_{\un{i} \un{j} })^{0\un{k}} {}_ {0\un{l}} \,
(E^{({\rm Ad})}_{\un{i} \un{j} })^{ 0 \un{l} } {}_ {0 \un{k} } ~,
\eea
where the following identity 
\be
(E^{({\rm Ad})}_{\un{i} \un{j} })^{\un{k}0} {}_ {\un{l}0} =
-(E^{({\rm Ad})}_{\un{i} \un{j} })^{0\un{l}} {}_ {0\un{k}} 
\ee
has been used. The group-theoretic factor in the last expression 
is also easy to evaluate:
\be 
\sum\limits_{\un{i} \neq \un{j}  }
 \sum\limits_{\un{k} ,\, \un{l}  } 
(E^{({\rm Ad})}_{\un{i} \un{j} })^{0\un{k}} {}_ {0\un{l}} \,
(E^{({\rm Ad})}_{\un{i} \un{j} })^{ 0 \un{l} } {}_ {0 \un{k} }
=(N-1)(N-2)~.
\label{GT-2}
\ee

As a result, the first and second  terms in  
(\ref{I+II-ad3}) lead  to the following contribution 
\be 
{\rm i} \,N(N-1) \,g^2\, {\bf G}^{(0)} (z,z') \Big\{ 
\hat{\bG}^{(e_{\rm a} )}\,\hat{\bG}'^{(e_{\rm a} )}
+ \hat{\bG}^{(-e_{\rm a} )}\,\hat{\bG}'^{(-e_{\rm a} )}\Big\}~.
\label{I+II-ad4}
\ee

We now turn to the third term in (\ref{I+II-ad3}). 
Since $\langle  v^{0 \un{i} }  \,v'^{ \un{i} 0  } \rangle
= {\rm i} \,g^2{\bf G}^{(e_{\rm a})} (z,z') $ 
is a massive propagator of charge $+ e_{\rm a}$, 
one of the hypermultiplet propagators
must be massive of charge $- e_{\rm a}$,
with the other must be free neutral. 
\bea
&& \qquad \qquad 
\sum\limits_{\un{i} } 
\langle  v^{0 \un{i} }  \,v'^{ \un{i} 0  }  \rangle  
 {\rm  tr}_{\rm Ad}  \left\{ E_{0 \un{i}  }
\hat{G} \,  E_{ \un{i} 0 }  \,\hat{G}'  \right\}   \\
&=& {\rm i} \,g^2\, {\bf G}^{(e_{\rm a})} (z,z') \,
\hat{\bG}^{(-e_{\rm a} )}\,\hat{\bG}'^{(0 )}
\Bigg\{ 
\sum\limits_{\un{i}, \, \un{j} }
\sum\limits_{I }
(E^{({\rm Ad})}_{0\un{i} })^I {}_ {0\un{j}} \,
(E^{({\rm Ad})}_{\un{i} 0 })^{ 0 \un{j} } {}_ {I } 
+ \sum\limits_{\un{i} ,\,\un{j} }
\sum\limits_{\un{k}\neq \un{l} }
(E^{({\rm Ad})}_{0\un{i} })^{\un{k}\un{l} } {}_ {0\un{j}} \,
(E^{({\rm Ad})}_{\un{i} 0 })^{ 0 \un{j} } {}_ {\un{k} \un{l} }
\Bigg\}    \non \\
&+& {\rm i} \,g^2\, {\bf G}^{(e_{\rm a})} (z,z') 
\hat{\bG}^{(0 )}\,
\hat{\bG}'^{(-e_{\rm a} )}
\Bigg\{ 
\sum\limits_{\un{i} ,\, \un{j}}
\sum\limits_{I }
(E^{({\rm Ad})}_{0\un{i} })^{0\un{j}}{}_I \,
(E^{({\rm Ad})}_{\un{i} 0 })^I{}_{ 0 \un{j} } 
+ \sum\limits_{\un{i},\, \un{j} }
\sum\limits_{\un{k}\neq \un{l} }
(E^{({\rm Ad})}_{0\un{i} })^{0\un{j}}{}_{\un{k}\un{l} } \,
(E^{({\rm Ad})}_{\un{i} 0 })^{\un{k}\un{l} }{}_{ 0 \un{j} } 
\Bigg\}~. \non 
\eea
Using the symmetry properties of the structure constants, 
the group-theoretical factors here can be related to those 
which occur in eqs. (\ref{GT-1}) and (\ref{GT-2}):
\bea
\sum\limits_{\un{i}, \, \un{j} }
\sum\limits_{I }
(E^{({\rm Ad})}_{0\un{i} })^I {}_ {0\un{j}} \,
(E^{({\rm Ad})}_{\un{i} 0 })^{ 0 \un{j} } {}_ {I } 
&=& 
\sum\limits_{\un{i} , \,\un{j}}
\sum\limits_{I }
(E^{({\rm Ad})}_{0\un{i} })^{0\un{j}}{}_I \,
(E^{({\rm Ad})}_{\un{i} 0 })^I{}_{ 0 \un{j} } 
=2(N-1)~,  \\
\sum\limits_{\un{i} ,\, \un{j} }
\sum\limits_{\un{k}\neq \un{l} }
(E^{({\rm Ad})}_{0\un{i} })^{\un{k}\un{l} } {}_ {0\un{j}} \,
(E^{({\rm Ad})}_{\un{i} 0 })^{ 0 \un{j} } {}_ {\un{k} \un{l} }
&=& 
\sum\limits_{\un{i} ,\,\un{j}}
\sum\limits_{\un{k}\neq \un{l} }
(E^{({\rm Ad})}_{0\un{i} })^{0\un{j}}{}_{\un{k}\un{l} } \,
(E^{({\rm Ad})}_{\un{i} 0 })^{\un{k}\un{l} }{}_{ 0 \un{j} } 
=(N-1)(N-2)~. \non 
\eea

As a result, the third  term in  
(\ref{I+II-ad3}) leads  to the following contribution 
\be 
{\rm i} \,N(N-1) \,g^2\, 
{\bf G}^{(e_{\rm a})} (z,z') \,
\Big\{ \hat{\bG}^{(-e_{\rm a} )}\,\hat{\bG}'^{(0 )}
+ \hat{\bG}^{(0 )}\,
\hat{\bG}'^{(e_{\rm a} )} \Big\} ~.
\label{I+II-ad5}
\ee

On the base of the above considerations, 
one can readily arrive at the final expression for
$\G_{\rm I+II}$:
\be
\G_{\rm I+II} = N(N-1)\,
\frac{g^2}{(4\p)^4} \, 
\U  \Big\{ {1\over 3} 
+  8 \, \hat{I}_{\rm I+II} \Big\} ~,
\ee
where 
\bea
 \hat{I}_{\rm I+II}  = \int\limits_{0}^{\infty} {\rm d}s_1 
{\rm d}s_2
{\rm d}s_3 \, 
{\rm e}^{-[ s_1 +  s_2]} \, 
\frac{ s_1^3 
[ s_1 + s_2 ]^2}
{[s_2s_3 +s_1s_2 +s_1 s_3]^{d/2 + 1} }
\label{I-I+II-ad}
\eea
is a divergent integral
in the limit $\ve = 4-d \to 0$. 
This integral follows from 
(\ref{I-I+II})
in the limit  $e_{\rm a} =e_{\rm f}=1$
or, equivalently, $N\to \infty$. 
Therefore, the divergent part of $\hat{I}_{\rm I+II}$  
can be read off from (\ref{I-I+II-div}),
\be
(\hat{I}_{\rm I+II})_{\rm div} = 
- {1 \over \ve} ~.
\ee

\subsection{Evaluation of $\G_{\rm III}$ }

The evaluation of $\G_{\rm III}$ is very similar 
to that of $\G_{\rm I+II}$ just described. 
Therefore, we simply give the final result: 
\be
\G_{\rm III} = {2 \over 3} \,N(N-1)\,
\frac{g^2}{(4\p)^4} \,\U ~.
\ee
No divergences are present. 

\subsection{Evaluation of $\G_{\rm IV}$ }

It remains to evaluate $\G_{\rm IV}$ which 
is determined by eq. (\ref{IV-mod}).
In  supersymmetric dimensional regularization, 
we have 
\bea
{1 \over 16} \, 
{\bar \cD}^2  \cD'^2 
\bG^{(e )}(z,z') \Big|_{z'=z} &=& 
\frac{(e^2 \, \f {\bar \f})^{d/2 -1} }{(4\p)^{d/2} }\,
\int\limits_{0}^{\infty}
\frac{ {\rm d} s }{s^{d/2} } \, {\rm e}^{-s}~, \qquad 
e \neq 0~, \non \\
{1 \over 16} \, 
{\bar \cD}^2  \cD'^2 
\bG^{(0 )}(z,z') \Big|_{z'=z} &=& 
0~.
\label{6-20}
\eea
The second relation here is actually a consequence 
of one of the fancy properties of 
dimensional regularization (see, e.g. \cite{Zinn})
\be
\int  \frac{ {\rm d}^d p }{p^2} = 0 \quad 
\Longleftrightarrow \quad 
\int\limits_{0}^{\infty}
\frac{ {\rm d} s }{s^{d/2}} =0~.
\label{SUSY-dim-reg}
\ee
Therefore, in the expression
\be
{\rm  tr}_{\rm Ad} \left\{ \{E_{0 \un{i} }, E_{\un{i} 0} \} \,
{\bar \cD}^2  \cD'^2 G^{({\rm Ad} )}(z,z') \Big|_{z'=z}
\right\}~,
\ee
which occurs in (\ref{IV-mod}), 
we should take into account the massive modes only. 
This amounts to computing the 
following group-theoretic factor
\be
\sum\limits_{\un{i} , \, \un{j} } \left(
\Big( \{E^{({\rm Ad})}_{0 \un{i} }, E^{({\rm Ad})}_{\un{i} 0} \} 
\Big)^{0 \un{j} }{}_{0 \un{j} }
+ \Big( \{E^{({\rm Ad})}_{0 \un{i} }, E^{({\rm Ad})}_{\un{i} 0} \} 
\Big)^{\un{j} 0}{}_{\un{j} 0} \right)= 2N(N-1)~.
\ee
As a result, we obtain 
\be
\G_{\rm IV} = -4 N(N-1)\,
\frac{g^2}{(4\p)^4} \, 
\U  \, I_{\rm IV}  ~,
\ee
with 
$ I_{\rm IV}  $ given in (\ref{IV-int}).
It is seen that the divergent parts of
of $  \G_{\rm I+II} $ and $ \G_{\rm IV} $ cancel each other, 
\be
\Big( \G_{\rm I+II}  +
\G_{\rm IV}  \Big)_{\rm div} = 0 ~,
\ee
consistent with the finiteness of the theory.
An alternative treatment of the cancellation of divergences
is given in the Appendix.

\sect{Discussion}
As pointed out in the Introduction, the two $\cN =2$ superconformal
field theories with gauge group $SU(N)$ considered 
in this paper differ only in the hypermultiplet sector 
--- one contains a single hypermultiplet in the
adjoint representation, the other contains $2N$ hypermultiplets 
in the fundamental representation. 
If the Dine-Seiberg conjecture holds,
then the two-loop $F^4$ contributions 
to the effective action must vanish in both theories. 
This would necessitate a cancellation 
of the $F^4$ corrections between the pure $\cN=2$ SYM, ghost 
and hypermultiplet sectors in both theories, implying that both 
theories should yield identical two-loop $F^4$ contributions 
in the hypermultiplet sector.

By explicit calculation, we have found  the following two-loop
$F^4$ contributions, $ \G_{\rm I+II} +  \G_{\rm III} +\G_{\rm IV} $,
from the hypermultiplet sector. For the case of
$\cN=2$ SYM with 
$2N$ hypermultiplets in the fundamental: 
\bea
\frac{g^2\, \U}{(4\p)^4} \,
\Big\{ N(N-1)(N-2) &+&N(N-1) \non \\
+  8N(N-1)  I_{\rm I+II} 
&-& 4(N-1)  I_{\rm III} 
-4(N-1)  (N -  2\,{N-1 \over N} )I_{\rm IV} \Big\} ~,
\label{EA1}
\eea
where the integrals $I_{\rm I+II},$ $I_{\rm III}$ and $I_{\rm IV}$
are given in equations (\ref{I-I+II}), (\ref{I-III}) and (\ref{IV-int})
respectively. For the case of $\cN=4$ SYM:
\be
\frac{g^2\,\U}{(4\p)^4} \,
\Big\{  N (N-1)
+  8 N (N-1) \, \hat{I}_{\rm I+II}  
-4  N(N-1) \,
  I_{\rm IV}\Big\} ~,
\label{EA2}
\ee
where the integrals $\hat{I}_{\rm I+II}$  and $I_{\rm IV}$
are given in equations (\ref{I-I+II-ad}) and (\ref{IV-int})
respectively.

In the large $N$ limit, all of the
integrals contained in the expressions
(\ref{EA1}) and (\ref{EA2})
 become independent 
of $N,$ as the charges $e_{\rm a}$ and $e_{\rm f}$
approach 1. 
With this observation, it is clear that these  
$F^4$ contributions have different large $N$ behaviour. 
The leading term in  (\ref{EA1}) 
is of order $N^3,$ while the leading term
in (\ref{EA2})  is of order $N^2.$
This is inconsistent with the Dine-Seiberg conjecture, which would
require identical leading large $N$ behaviour for the two theories.

It is instructive to examine the source of the difference
in the large $N$
behaviour of the two theories, 
which is due to the presence of a
 leading $N^3$ contribution in the case 
of $\cN=2$ SYM with $2N$
hypermultiplets in the fundamental. This contribution
 comes from  the diagrams of type I, II and III
in which the $\cN=2$ vector multiplet 
propagator (that is, $\langle v\, v \rangle$ 
or $\langle \vf \, \vf^\dagger \rangle$)
is massless and corresponds to one of the 
unbroken $SU(N-1)$ generators $E_{\un{i} \un{j} }$,
and the two hypermultiplet propagators are massive
with the same mass $\bar{\phi} \phi/N(N-1)$.
In the large $N$ limit, these
hypermultiplets become massless
and decouple from the background (as
their $U(1)$ charges, $\pm 1/ \sqrt{ N(N-1))}$,  
vanish), and so
one might at first sight expect these diagrams not to
contribute terms proportional to $\Upsilon$ in the
large $N$ limit. However, the situation is more subtle, 
because they really decouple only for $N= \infty$.
The point is that the magnitude of the
$U(1)$ charge on each of the hypermultiplet lines is the same. 
Since all charge dependence occurs in the form $e W$ or $e \phi,$ 
it cancels out of  the terms proportional to
$W^2 \bar{W}^2/ (\f \bar{\phi} )^2$, see eqs. 
(\ref{FUN-1}) and (\ref{FUN-2}).
As a result, the contribution from these diagrams
 survives in the large $N$ limit. 
There is a combinatoric factor of $2N (N-1) (N-2),$ as
there are $2N$ hypermultiplets and 
$(N-1)(N-2)$ massless vectors
 $v^{\un{i} \un{j}}.$

There remains a (pretty solid)
hope that the Dine-Seiberg conjecture
holds, at least in the large $N$ limit, for those $\cN=2$ 
superconformal theories which possess supergravity duals, 
in particular: (i) $\cN=4$ SYM; 
(ii) $USp(2N)$ gauge theory  with a traceless antisymmetric
hypermultiplet and four fundamental hypermultiplets
\cite{ASST}; (iii) quiver gauge theories \cite{quiver}. 
This is based upon the AdS/CFT correspondence. 
Maximal supersymmetry should also play a crucial role 
in the case of $\cN=4$ SYM.
Otherwise one would be forced to re-consider 
numerous conclusions drawn on the basis of 
this conjecture, for instance, in \cite{BPT,KMT}. 
Explicit two-loop calculations of $F^4$ corrections
in such theories are therefore 
extremely desirable and can be 
carried out using the techniques developed in the present 
paper in conjunction with some ideas given in \cite{KMT}. 

\vskip.5cm

\noindent
{\bf Acknowledgements:}\\
We are grateful to Joseph Buchbinder, Jim Gates
and Arkady Tseytlin for  comments.
This work is supported in part by the Australian Research
Council and  UWA research grants.

\begin{appendix}

\sect{Cancellation of divergences}
To handle ill-defined two-loop  integrals, 
we employed supersymmetric regularization 
via dimensional reduction. Its use allowed us, 
in a safe yet simple way, to make sure that 
no divergences are present. On the other hand, 
the absence of divergences indicates that there
should exist a manifestly finite form for the effective 
action  directly in four space-time dimensions.  
Here we elaborate on such a form 
in the case of $\cN=4$ SYM. 

The second and third terms in the two-loop 
contribution (\ref{EA2}) contain the  
proper-time integrals $\hat{I}_{\rm I+II} $
(\ref{I-I+II-ad}) and $ I_{\rm IV} $  (\ref{IV-int}), 
each of which diverges in $d=4$. 
Nevertheless, let us try to evaluate 
the joint contribution coming  from 
the second and third terms in  (\ref{EA2})
in $d=4$.  Since we no longer use 
supersymmetric dimensional regularization, 
the integral  $ I_{\rm IV} $ has to be
modified as follows
\be
I_{\rm IV} \quad \longrightarrow \quad
\hat{I}_{\rm IV}  = \int\limits_{0}^{\infty} 
{ {\rm d}s \over s^{d/2} } \,  {\rm e}^{-s} 
+\int\limits_{0}^{\infty} 
{ {\rm d}s \over s^{d/2} }~. 
\label{IV-int-til}
\ee
The second term here 
is generated by those supergraphs
of type IV
which involve the structure in the second line 
of (\ref{6-20}). This term 
cannot be ignored anymore, 
since the identity (\ref{SUSY-dim-reg})
holds only in the framework of 
supersymmetric dimensional regularization.
The sum of divergent integrals is 
\bea 
\Big( 2\hat{I}_{\rm I+II} - 
\hat{I}_{\rm IV} \Big)\Big|_{d=4} 
= 2\int\limits_{0}^{\infty} {\rm d}s \,
{\rm d}t \,
{\rm d}u \, 
\frac{ t^3 \,
( s + t )^2}
{(st  +su +tu)^{3 }}\,
{\rm e}^{-( s +  t)} 
-  \int\limits_{0}^{\infty} 
{ {\rm d}s \over s^2 }
\,  {\rm e}^{-s} 
-  \int\limits_{0}^{\infty} 
{ {\rm d}s \over s^2 }
~.
\eea
In the first term on the right, 
one can easily do the $u$-integral:
\bea
\int\limits_{0}^{\infty} {\rm d}s \,
{\rm d}t \,
{\rm d}u \, 
\frac{ t^3 \,
( s + t )^2}
{(st  +su +tu)^{3 }}\,
{\rm e}^{-( s +  t)} 
&=& \hf   \int\limits_{0}^{\infty} {\rm d}s \,
{\rm d}t \,
\frac{ t \,
( s + t )}
{s ^{2 }}\,
{\rm e}^{-( s +  t)} \non \\
= \hf   \int\limits_{0}^{\infty} {\rm d}t \,
t\,{\rm e}^{-t} 
\int\limits_{0}^{\infty} { {\rm d}s \over s} \,  
{\rm e}^{-s} 
&+& 
\hf   \int\limits_{0}^{\infty} {\rm d}t \,
t^2 \,{\rm e}^{-t} 
\int\limits_{0}^{\infty} { {\rm d}s \over s^2} \,  
{\rm e}^{-s} ~.
\eea
As a result, one gets 
\bea 
\Big( 2\hat{I}_{\rm I+II} - \hat{I}_{\rm IV} \Big)\Big|_{d=4} 
= \int\limits_{0}^{\infty}  {\rm d}s \,
{ {\rm d} \over {\rm d}  s} \,
\left\{
\frac{1}{s} \,  ( 1 - {\rm e}^{-s} ) \right\}
= -1~.
\eea
This shows that the second and third terms 
in  (\ref{EA2}) provide a finite contribution
to the effective action.

\end{appendix}


\begin{thebibliography}{99}

\bibitem{KM}
S.~M.~Kuzenko and I.~N.~McArthur,
``On the background field method beyond one loop: 
A manifestly covariant  derivative expansion 
in super Yang-Mills theories,''
JHEP {\bf 0305} (2003) 015 [arXiv:hep-th/0302205].

\bibitem{KM2}
S.~M.~Kuzenko and I.~N.~McArthur, 
``Low-energy dynamics in N = 2 super QED: Two-loop approximation,''
JHEP {\bf 0310} (2003) 029
[arXiv:hep-th/0308136].

\bibitem{BKO}
I.~L.~Buchbinder, S.~M.~Kuzenko and B.~A.~Ovrut,
``On the D = 4, N = 2 non-renormalization theorem,''
Phys.\ Lett.\ B {\bf 433} (1998) 335
[arXiv:hep-th/9710142].

\bibitem{BBKO}
I.~L.~Buchbinder, E.~I.~Buchbinder, S.~M.~Kuzenko and B.~A.~Ovrut,
``The background field method for N = 2 super Yang-Mills theories 
in  harmonic superspace,''
Phys.\ Lett.\ B {\bf 417} (1998) 61
[arXiv:hep-th/9704214].

\bibitem{DS}
M.~Dine and N.~Seiberg,
``Comments on higher derivative operators in some 
SUSY field theories,''
Phys.\ Lett.\ B {\bf 409} (1997) 239
[arXiv:hep-th/9705057].

\bibitem{dWGR}
B.~de Wit, M.~T.~Grisaru and M.~Ro\v{c}ek,
``Nonholomorphic corrections to the one-loop 
N=2 super Yang-Mills action,''
Phys.\ Lett.\ B {\bf 374} (1996) 297
[arXiv:hep-th/9601115].

\bibitem{Hen}
M.~Henningson,
``Extended superspace, higher derivatives 
and SL(2,Z) duality,''
Nucl.\ Phys.\ B {\bf 458} (1996) 445
[arXiv:hep-th/9507135].

\bibitem{BKT}
I.~L.~Buchbinder, S.~M.~Kuzenko and A.~A.~Tseytlin,
``On low-energy effective actions in N = 2,4 superconformal theories 
in  four dimensions,''
Phys.\ Rev.\ D {\bf 62} (2000) 045001
[arXiv:hep-th/9911221].

\bibitem{GR}
F.~Gonzalez-Rey and M.~Ro\v{c}ek,
``Nonholomorphic N = 2 terms in N = 4 SYM: 
1-loop calculation in N = 2  superspace,''
Phys.\ Lett.\ B {\bf 434} (1998) 303
[arXiv:hep-th/9804010];
F.~Gonzalez-Rey, B.~Kulik, I.~Y.~Park and M.~Ro\v{c}ek,
``Self-dual effective action of N = 4 super-Yang-Mills,''
Nucl.\ Phys.\ B {\bf 544} (1999) 218
[arXiv:hep-th/9810152].

\bibitem{BK2}
I.~L.~Buchbinder and S.~M.~Kuzenko,
``Comments on the background field method 
in harmonic superspace:  
Non-holomorphic corrections in N = 4 SYM,''
Mod.\ Phys.\ Lett.\ A {\bf 13} (1998) 1623
[arXiv:hep-th/9804168];
E.~I.~Buchbinder, I.~L.~Buchbinder and S.~M.~Kuzenko,
``Non-holomorphic effective potential in N = 4 SU(n) 
SYM,'' Phys.\ Lett.\ B {\bf 446} (1999) 216
[arXiv:hep-th/9810239].

\bibitem{LvU}
D.~A.~Lowe and R.~von Unge,
``Constraints on higher derivative operators 
in maximally supersymmetric  gauge theory,''
JHEP {\bf 9811} (1998) 014
[arXiv:hep-th/9811017].

\bibitem{DKMSW}
N.~Dorey, V.~V.~Khoze, M.~P.~Mattis, M.~J.~Slater and W.~A.~Weir,
``Instantons, higher-derivative terms, 
and nonrenormalization theorems 
in  supersymmetric gauge theories,''
Phys.\ Lett.\ B {\bf 408} (1997) 213
[arXiv:hep-th/9706007].

\bibitem{BFMT}
D.~Bellisai, F.~Fucito, M.~Matone and G.~Travaglini,
``Non-holomorphic terms in N = 2 SUSY 
Wilsonian actions and RG equation,''
Phys.\ Rev.\ D {\bf 56} (1997) 5218
[arXiv:hep-th/9706099].

\bibitem{BK} I.~L.~Buchbinder and S.~M.~Kuzenko,
{\it Ideas and Methods of Supersymmetry and
Supergravity or a Walk Through Superspace},
IOP, Bristol, 1998.

\bibitem{GGRS}
S.~J.~Gates, M.~T.~Grisaru, M.~Ro\v{c}ek and W.~Siegel,
{\it Superspace, or One Thousand 
and One Lessons in Supersymmetry},
Benjamin/Cummings, 1983 [arXiv:hep-th/0108200].

\bibitem{OW}
B.~A.~Ovrut and J.~Wess,
``Supersymmetric $R_\x$ gauge and radiative symmetry breaking,''
Phys.\ Rev.\ D {\bf 25} (1982) 409;
P.~Binetruy, P.~Sorba and R.~Stora,
``Supersymmetric S covariant $R_\x$ gauge,''
Phys.\ Lett.\ B {\bf 129} (1983) 85.

\bibitem{BBP}
A.~T.~Banin, I.~L.~Buchbinder and N.~G.~Pletnev,
``On low-energy effective action in N = 2 super 
Yang-Mills theories on  non-abelian background,''
Phys.\ Rev.\ D {\bf 66} (2002) 045021
[arXiv:hep-th/0205034];
``One-loop effective action for N = 4 SYM theory 
in the hypermultiplet  sector: Leading low-energy 
approximation and beyond,''
Phys.\ Rev.\ D {\bf 68} (2003) 065024
[arXiv:hep-th/0304046].

\bi{Georgi} H. Georgi, {\it Lie Algebras in Particle Physics: From 
Isospin to Unified Theories}, Benjamin/Cummings, 1982.

\bi{Zinn} J. Zinn-Justin, {\it Quantum Field Theory and 
Critical Phenomena}, Oxford University Press, 1989.

\bibitem{ASST}
O.~Aharony, J.~Sonnenschein, S.~Yankielowicz and S.~Theisen,
``Field theory questions for string theory answers,''
Nucl.\ Phys.\ B {\bf 493} (1997) 177
[arXiv:hep-th/9611222]; 
M.~R.~Douglas, D.~A.~Lowe and J.~H.~Schwarz,
``Probing F-theory with multiple branes,''
Phys.\ Lett.\ B {\bf 394} (1997) 297
[arXiv:hep-th/9612062];
O.~Aharony, J.~Pawelczyk, S.~Theisen and S.~Yankielowicz,
``A note on anomalies in the AdS/CFT correspondence,''
Phys.\ Rev.\ D {\bf 60} (1999) 066001
[arXiv:hep-th/9901134].

\bibitem{quiver}
M.~R.~Douglas and G.~W.~Moore,
``D-branes, quivers, and ALE instantons,''
arXiv:hep-th/9603167; 
C.~V.~Johnson and R.~C.~Myers,
``Aspects of type IIB theory on ALE spaces,''
Phys.\ Rev.\ D {\bf 55} (1997) 6382
[arXiv:hep-th/9610140];
S.~Kachru and E.~Silverstein,
``4d conformal theories and strings on orbifolds,''
Phys.\ Rev.\ Lett.\  {\bf 80} (1998) 4855
[arXiv:hep-th/9802183].

\bibitem{BPT}
I.~L.~Buchbinder, A.~Y.~Petrov and A.~A.~Tseytlin,
``Two-loop N = 4 super Yang Mills effective action 
and interaction  between D3-branes,''
Nucl.\ Phys.\ B {\bf 621} (2002) 179
[arXiv:hep-th/0110173].

\bibitem{KMT}
S.~M.~Kuzenko, I.~N.~McArthur and S.~Theisen,
``Low energy dynamics from deformed conformal symmetry 
in quantum 4D N = 2 SCFTs,''
Nucl.\ Phys.\ B {\bf 660} (2003) 131
[arXiv:hep-th/0210007].

\end{thebibliography}
\end{document}